\pdfoutput=1 

\documentclass[sigconf, balance, nonacm]{acmart}

\usepackage{hologo}
\usepackage{xspace}
\usepackage[ruled,linesnumbered]{algorithm2e}
\SetKw{Continue}{continue}
\usepackage{pifont}
\usepackage{CJKutf8}
\usepackage{makecell}
\usepackage{multirow}
\usepackage{multicol}
\usepackage{tablefootnote}
\usepackage{CJKutf8}
\usepackage{xcolor}
\usepackage[T1]{fontenc}
\usepackage[utf8]{inputenc}
\usepackage{url}
\usepackage{wasysym}
\usepackage{threeparttable}
\usepackage{marvosym}

\newif\ifSimpleMode\SimpleModefalse

\newcounter{ToDo}

\newcommand{\SysName}{\textsc{Marlin}\xspace}

\newcommand{\CADETS}{{\scshape Cadets} }

\newcommand{\THEIA}{{\scshape Theia} }

\AtBeginDocument{%
  \providecommand\BibTeX{{%
    \normalfont B\kern-0.5em{\scshape i\kern-0.25em b}\kern-0.8em\TeX}}}

\begin{document}

\title{\SysName: Knowledge-Driven Analysis of Provenance Graphs for Efficient and Robust Detection of Cyber Attacks}



\author{Zhenyuan Li*, Yangyang Wei*, Xiangmin Shen\textsuperscript{\dag}, Lingzhi Wang\textsuperscript{\dag}, Yan Chen\textsuperscript{\dag}, Haitao Xu*, Shouling Ji*, Fan Zhang*, Liang Hou\textsuperscript{\P}, Wenmao Liu\textsuperscript{\ddag}, Xuhong Zhang*, Jianwei Ying*}
\affiliation{%
  \institution{*Zhejiang University, \textsuperscript{\dag}Northwestern University, \textsuperscript{\P}GuoTai Junan Security, \textsuperscript{\ddag}NSFOCUS Technologies Group Co., Ltd.}
  \country{}
}



\begin{abstract}
Recent research in both academia and industry has validated the effectiveness of provenance graph-based detection for advanced cyber attack detection and investigation. 
However, analyzing large-scale provenance graphs often results in substantial overhead. 
To improve performance, existing detection systems implement various optimization strategies. 
Yet, as several recent studies suggest, these strategies could lose necessary context information and be vulnerable to evasions. 
Designing a detection system that is efficient and robust against adversarial attacks is an open problem. 

We introduce \SysName, which approaches cyber attack detection through real-time provenance graph alignment.
By leveraging query graphs embedded with attack knowledge, \SysName can efficiently identify entities and events within provenance graphs, embedding targeted analysis and significantly narrowing the search space. 
Moreover, we incorporate our graph alignment algorithm into a tag propagation-based schema to eliminate the need for storing and reprocessing raw logs. 
This design significantly reduces in-memory storage requirements and minimizes data processing overhead. 
As a result, it enables real-time graph alignment while preserving essential context information, thereby enhancing the robustness of cyber attack detection.
Moreover, \SysName allows analysts to customize attack query graphs flexibly to detect extended attacks and provide interpretable detection results.

We conduct experimental evaluations on two large-scale public datasets containing 257.42 GB of logs and 12 query graphs of varying sizes, covering multiple attack techniques and scenarios.
The results show that \SysName can process 137K events per second while accurately identifying 120 subgraphs with 31 confirmed attacks, along with only 1 false positive, demonstrating its efficiency and accuracy in handling massive data.
Moreover, \SysName successfully detected all 3 evasion attacks that typical existing systems would miss, demonstrating its robustness against adversarial attacks.
\end{abstract}

\maketitle
\section{Introduction}
\label{sec:intro}

Recently, numerous cyber attacks targeted essential infrastructures globally, including power grids, water treatment facilities, and financial institutions, causing significant damage~\cite{infrastructure}. 
In response to these sophisticated and escalating threats, the DARPA Transparent Computing project proposes a detection strategy emphasizing high-fidelity monitoring of system activities with provenance graphs (aka causality graphs) combined with a comprehensive analysis of global behavior~\cite{TC}.
Recent concentrate research on provenance graph-based intrusion detection (PIDSes)~\cite{datta2022alastor, rehman2024flash, li2023nodlink, li2019effective, munoz2017efficient} has suggested its success. 

Nevertheless, high-fidelity and comprehensive observations of system behaviors produce vast data and require substantial analytical resources.
To manage computational complexity, security participants usually apply performance-optimizing strategies to decrease the size of actually analyzed data or limit analysis scopes to improve performance. 
Commonly used strategies include anomaly-based path filtering, which analyzes only provenance paths exhibiting out-of-baseline behavior~\cite{nodoze, provdetector}.
Another approach involves narrowing the analysis scope by restricting the tag propagation time and rounds or the size of the graph embedded for GNN-based approaches, which helps to avoid the overwhelming computational load caused by extensive dependency explosion~\cite{rehman2024flash, hossain2020combating}.

Recent research~\cite{goyal2023sometimes, mukherjee2023evading} has revealed vulnerabilities in these methods. 
Attackers could utilize these strategies to evade PIDSes by altering the provenance graph and inserting benign events into or around attack chains.
Techniques such as control flow hijacking allow attackers to modify the provenance graph during real-world attacks. 
Another typical evasion strategy targets rule-based detection methods by constructing mutated attacks~\cite{li2022attackg}, including the modification of attack entities, and combinations of these techniques.
Table~\ref{tab:acomparsion} compares the robustness of seven state-of-the-art PIDSes against these evasion methods, including \SysName. 
While GNN-based and anomaly detection systems excel at identifying mutated attacks, they often have a limited information aggregation scope and could be affected by benign activities around attack chains, making them susceptible to benign event insertion attacks. In contrast, rule-based systems can analyze longer attack chains, are harder to evade. However, they tend to be less effective at detecting mutated attacks.
Thus, developing a provenance-based detection system that is both efficient and robust against aforementioned adversarial attacks is still a significant open research problem.


\begin{table}[htbp]
    \footnotesize
        \centering
        \vspace{-0.1in}
        \caption{Robustness of PIDSes against Adversarial Attacks}
        \vspace{-0.1in}
        \begin{tabular}{c|c|c|c}
        \toprule 
            \multicolumn{1}{c|}{Approaches} & \makecell{In-chain Insertion\\ProvNinja.~\cite{mukherjee2023evading}} & \makecell{Around Insertion\\Goyal et al.~\cite{goyal2023sometimes}} & \makecell{Mutated-\\attack} \\
        \midrule
              \textsc{Streamspot}~\cite{StreamSpot}  &\ding{55}  &\ding{55}  &\ding{51}   \\ 
              \textsc{Unicorn}~\cite{han2020unicorn}             & \ding{55} &\ding{55}  &\ding{51}   \\
              \textsc{NodLink}~\cite{li2023nodlink}          & \ding{55} &\ding{55}  &\ding{51}   \\
              \textsc{Flash}~\cite{rehman2024flash}  &\ding{55}  &\ding{51}  &\ding{51}   \\ 
              \textsc{Morse}~\cite{morse}        &\ding{51}  &\ding{51}  &\ding{55}   \\
                \SysName (our work)     &\ding{51} & \ding{51} & \ding{51}  \\
        \bottomrule
        \end{tabular}
        \vspace{-0.1in}
        \label{tab:acomparsion}
\end{table}


Toward this target, we proposed \SysName in this paper, which conceptualizes real-time attack detection in streaming logs as a streaming graph alignment problem. 
Graph querying and matching tasks typically require significant computational effort due to the necessity of traversing numerous branches to identify potential matches~\cite{livi2013graph}.
Nonetheless, using query graphs annotated with seeds where searches start, we can identify key entities and events within the provenance graph, significantly narrowing the search space and enabling continuous and detailed analysis of streaming provenance graphs with minimal computational overhead. 
Additionally, fuzzified node and edge matching conditions are employed to ensure comprehensive coverage of mutated attacks. 
These query graphs can either represent the whole attack for accurate attack location or attack techniques for better coverage of various attack technique combinations.
Subsequently, we exploit the inherent causality in provenance graphs and incorporate the graph alignment process into the tag propagation workflow. 
This workflow includes tag initialization (waiting for seeds), tag propagation (extending the search), alert triggering (confirming alignment), and tag removal (ending the search).
This integrated approach significantly reduces computational overhead by processing each event only once, eliminating the need to cache the entire graph and thereby reducing memory usage.

To evaluate \SysName's accuracy and efficiency in attack subgraph alignment, we comparative evaluated it on two publicly available datasets, namely, the DARPA T3 Engagement 3 dataset~\cite{TC} and the PKU-ASAL dataset~\cite{asal}, which comprise a total of 257.42GB of data collected over 694 hours and dozens of attack stages. 
Concurrently, we construct 12 query graphs, 6 of which correspond to attack techniques or simple combinations, while the remaining 6 correspond to complete attack scenarios, typically larger in scale.
As shown, \SysName identifies all 120 aligned subgraphs and 31 confirmed attacks with just 1 false positive, achieving recall and precision rates of 1.0 and 0.96, respectively. Moreover, it successfully detected all 3 evasion attacks, demonstrating its outperforming robustness. In terms of efficiency, \SysName can process 137K events per second with only 6 CPU cores and less than 400MB of RAM, which means that \SysName running on a personal computer can process logs from approximately 6,000 endpoints.
In summary, this paper makes the following contributions:

\vspace{-\topsep}
\begin{list}{\labelitemi}{\leftmargin=1.5em}
 \setlength{\topmargin}{0pt}
 \setlength{\itemsep}{0em}
 \setlength{\parskip}{0pt}
 \setlength{\parsep}{0pt}
    \item We conceptualize real-time attack detection in streaming logs as a streaming graph alignment problem. Utilizing the attack knowledge encoded in query graphs, we are able to accurately and efficiently pinpoint key entities and events within the provenance graph and narrow down the search space while maintaining necessary contexts to be robust against evasions.
    
    \item Leveraging the causality in provenance graphs, we integrate the streaming graph alignment process into a tag propagation framework and present \SysName. With carefully designed tag cache and propagation rules, each event is processed only once, eliminating the need for repeated caching and processing of raw logs, making \SysName an efficient detection system. 

    \item We conducted a systematic evaluation of \SysName on two open-source, large-scale datasets. The results show that \SysName not only achieves superior accuracy, with recall and precision rates of 1.0 and 0.96, respectively but also demonstrates robustness against evasion attacks. Moreover, it maintains high efficiency, processing streaming events at a throughput of 13.7K events per second.

\vspace{-\topsep}
\end{list}
\section{Background}
\label{sec:background}

\subsection{Provenance Graph-based Attack Detection}
\label{subsec:pgdetection}

The concept of provenance originates from data lineage analysis.~\cite{ikeda2009data} 
In intrusion detection, provenance analysis aims to identify potential attack paths by tracing data and control flow.
Typically, provenance analysis is performed on entities in the system, such as process, files, and their interactions with each other, such as reading files, and forking processes. 
Various data collection tools for different operating systems have been proposed for provenance analysis utilizing mechanisms such as eBPF~\cite{ebpf} and ETW~\cite{etw}.
Typically, these tools will fetch audit log data actively and parse them into a unified 4-tuple $(s,v,o,t)$ event format, in which $s$ denotes subjects of operations, $v$ denotes operation details, $o$ denotes objects of behaviors and $t$ denotes occurring time of operations.
These 4-tuple form temporal graphs through shared nodes, including subjects and objects.
Several recent studies~\cite{chen2021clarion, datta2022alastor, hassan2020omegalog} have expanded the scope of research by incorporating additional data sources, such as serverless functions and web applications, to enhance observability.
Meanwhile, a variety of provenance-graph-based detection schemes~\cite{altinisik2023provg, li2021threat, li2023nodlink} have been proposed, demonstrating that provenance analysis excels in attack semantic analysis, correlation behavior analysis, and interpretability of detection results.

Despite the fact that provenance graphs offer substantial benefits in enhancing the efficiency and accuracy of intrusion detection systems, they also pose extensive computational and storage challenges, particularly in real-time scenarios. Due to performance constraints, PIDSs adopt various performance-enhancing strategies.
Machine learning-based approaches typically detect anomalies by learning patterns of benign behavior within provenance graphs. Systems utilizing Graph Neural Networks to identify potential threats through deviations from these learned patterns. The capability of GNNs largely depends on their architecture, specifically the breadth (size of the embedded graph) and depth (number of layers). For extensive provenance graphs, adding layers can exponentially increase the number of interconnected nodes, complicating the learning of node relationships. To manage this complexity, techniques like subgraph-based sampling~\cite{zeng2019graphsaint, chiang2019cluster} are employed to constrain the receptive field of the GNN, thereby narrowing the analytical scope. Besides, higher-level nodes may lose information after multiple rounds of message passing, which can lead to a decline in the network's representational power. Rule-based methods~\cite{holmes, morse} also limit the depth and scope of tag propagation or use path filtering to reduce the volume of data actually analyzed.
While these strategies effectively reduce computational overhead, they also involve potential vulnerabilities that attackers could exploit.

\subsection{Evasion Against Provenance Graph-based Attack Detection}
\label{subsec:evasion}


\begin{figure}[htbp]
    \footnotesize
	\centering
     \vspace{-0.1in}
	\includegraphics[width=0.98\linewidth]{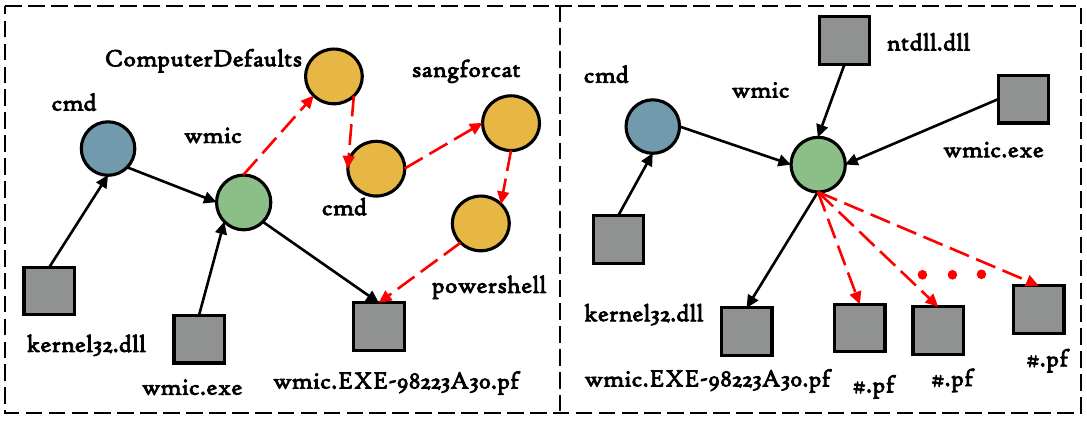}
     \vspace{-0.1in}
     \caption{Examples of Evasion Attack by In-chain Insertion (Left) and Around Insertion (Right)}
    \label{fig: mimic example}
    \vspace{-0.1in}
\end{figure}

Towards the aforementioned vulnerabilities, several methods have been proposed to evade existing Provenance-based Intrusion Detection Systems (PIDSes). 
Goyal et al.~\cite{goyal2023sometimes} suggest that adversaries can alter graph embeddings by introducing additional benign activities or by incorporating benign structures that filter out malicious ones during downsampling, thus impairing the system's detection capabilities. 
Subsequently, Mukherjee et al.~\cite{mukherjee2023evading} introduce the concept of a "gadget chain," a sequence of events designed to replace more conspicuous ones. This strategy minimizes the use of rare events and effectively achieves evasion by reducing the saliency of attacks.
It is noteworthy that although utilizing these bypass methods in real-world attacks is not always easy, a successful implementation can almost guarantee complete circumvention of the detection architecture, making it difficult for defenders to counter these attacks with minor improvements.

Fig.~\ref{fig: mimic example} illustrates two evasion attack examples. Firstly, evasion by in-chain insertion where attackers replace the suspicious behavior of \texttt{wmic} operating \texttt{WMIC.EXE-98223A30.pf} with a gadget chain. This gadget chain maintains the same start and end points as the original behavior to ensure functional consistency, but it intersperses multiple common behaviors to obscure the original suspicious activity. This strategy can  evades detection methods with a limited analysis scope.
Secondly, evasion by around-chain insertion, where adversaries add modification operations to similar files like \texttt{*.pf} around the \texttt{wmic} process, aiming to make the nodes in the attack graph have similar embeddings as the nodes of benign activities, thereby bypassing mostly GNN-based detection.

Another typical evasion strategy targets rule-based detection methods by constructing mutated attacks~\cite{li2022attackg} (Detailed discussed with an example in Appendix~\S\ref{app:mutation}). These attacks involve detailed implementation of attack techniques, selection of attack nodes, and combinations of these techniques. To counteract these, we can employ anomaly detection principles or enhance detection rules to offer broader generalization, adapting to novel attacks. In this paper, the query graphs used in this paper can be constructed at the technique level to counter new combinations of attack techniques. They also integrate various attack node selections and methods to maximize coverage of potential attack implementations.

\subsection{Cyber Attack Knowledge Representation}

Recently, Cyber Threat Intelligence (CTI) embedded with attack knowledge is widely used by security analysts to counter evolving cyber-attacks. To manage and communicate CTI knowledge more effectively, the security community has widely adopted standardized open formats such as OpenIoC~\cite{OpenIoC}, STIX~\cite{STIX}, and CybOX~\cite{Cybox} to describe entity-level indicators of compromise (IoCs) accurately.

Recently, graph-level attack representation has gained wide attention for better expressiveness and robustness. 
Multiple tools, such as Extractor~\cite{satvat2021extractor}, AttacKG~\cite{li2022attackg}, and others~\cite{gao2021enabling, legoy2020automated, legoy2020automated}, have been proposed to construct graph-based cyber attack representation. Specifically, AttacKG has made significant progress by constructing technical knowledge graphs (TKGs) at the attack technique level and merging knowledge from multiple sources to gain a more robust representation against attack mutations. 
\SysName utilizes these query graphs to precisely locate attacks within the provenance graph, effectively narrowing the search space. This paper specifically focuses on optimizing graph alignment to be both lightweight and efficient, leveraging the tag propagation framework.


\section{System Overview}
\label{sec:overview}

\begin{figure}[htbp]
    \centering
    \includegraphics[width=0.47\textwidth]{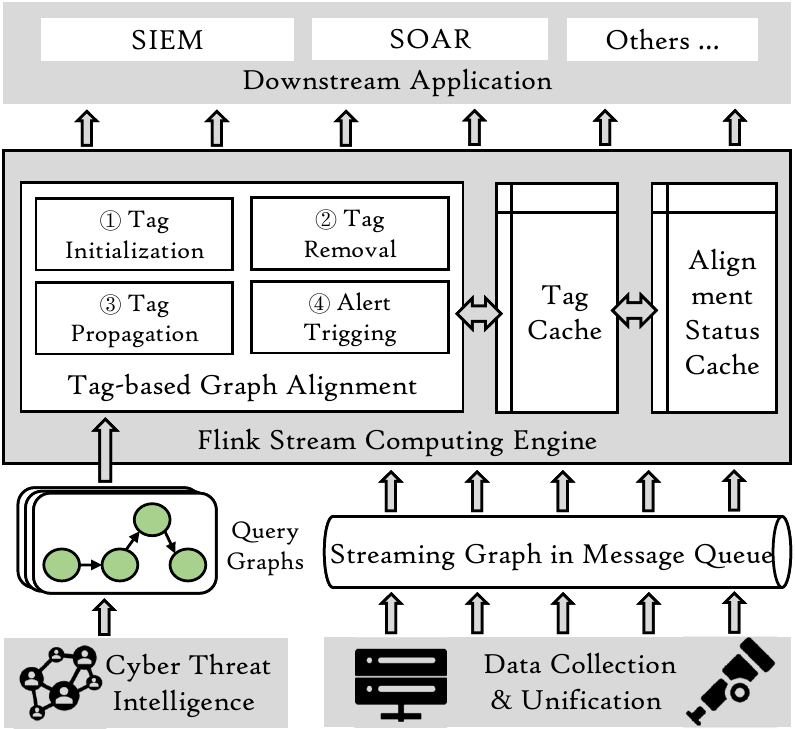}
    \vspace{-0.05in}
    \caption{Architecture of \SysName}
    \vspace{-0.2in}
    \label{fig:system_overview}
\end{figure}

Fig.~\ref{fig:system_overview} presents the overall architecture of \SysName. 
The system comprises three primary components: the data collection module, the query graph generation module, and the tag propagation-based detection module. 
Given that the first two modules are based on more established solutions, this paper primarily focuses on the intricacies and advancements within the third module.

\subsection{Streaming Provenance Graph Processing}

\SysName is designed to align streaming provenance graphs and is compatible with all operating systems. It can integrate data from various sources, such as ETW and eBPF, standardizing it into a uniform format for analysis.
In operation, \SysName deals with the continuous, real-time generation of massive events, referred to as the streaming provenance graph. The rapid generation of this data stream poses significant challenges, as even short-term data caching for analysis requires substantial memory and incurs significant query overhead. Therefore, it is essential to develop a framework and algorithm for streaming provenance graph analysis that minimizes cache size without compromising the advantages of provenance analysis.

In this paper, we use tags to cache intermediate results, avoid repeated consumption of events, and directly cache original logs.
Besides, we design tag propagation rules, which define strategies for initializing tags, where to propagate them, and cache updates to ensure that each event needs to be processed only once.
Intuitively, tags cache the result of precedence behavior summarized by propagation rules and can be viewed as the state of a particular entity at a specific time.
Thus, We can utilize an entity-to-tag map to cache tags in memory. 
It is worth noting that this approach is more efficient than caching the entire graph ($O(N)$ vs $O(N^2)$), especially for provenance graphs of long-running systems where the number of events exceeds the number of entities. 

\subsection{Generating Query Graph}

In this paper, we model the real-time cyber attack detection problem as a streaming graph alignment problem, where query graphs contain knowledge of attacks and are generalized to cover similar attacks.
Recent works~\cite{li2022attackg, husari2017ttpdrill, satvat2021extractor} have shown the potential of using NLP technologies to automatically extract threat knowledge and aggregate it into graph-structured attack patterns from threat intelligence reports. 
We adopt a hybrid approach, combining both automated and manual methods, to optimize previous efforts in producing attack graphs and achieve better detection results with sparse attack descriptions.

Furthermore, we suggest constructing query graphs at the granularity of attack techniques to manage their size and boost detection efficiency. This granularity can also enhance the robustness of detection by covering the varied combinations of attack techniques.
To further broaden the scope of attacks that the query graphs can detect, we implement generalization strategies. These include integrating different implementations of attack techniques within the same query graph and using regular expressions (like ``\texttt{*(tar|7z|winrar|)*}'' for compression-related operations) to describe node characteristics. We set specific alignment thresholds, considering an alignment successful when the matched content meets or exceeds these thresholds.
In this paper, we provide 12 query graphs covering multiple attack techniques and scenarios, detailed in Appendix~\ref{app:query_graphs}.
In practice, users can customize query graphs to meet varying targets.

\subsection{Tag Propagation-based Graph Alignment}

Using tag-propagation for provenance graph analysis is not a new idea. 
Some studies have attempted to use tags to analyze sensitive data leakage and the combination of attack stages in systems. 
However, the structure and propagation policies of these tags are relatively simple. 
It is still an open question whether more complex tags can be designed to handle a more complex problem like graph alignment on stream while maintaining high efficiency.

Specifically, there are several challenges to overcome. 
First, we need to try to aggregate the alignment results of multiple branches in the graph with a single-line tag propagation.
In this paper, we adopt a ``Tag - Alignment Status'' two-tier caching structure to merge aligned results in the second caching.
It is worth noting that the two-tier cache we use already merges multiple branches. Therefore, no additional alert merging scheme is required. 

Then, we adopt several strategies to minimize the processing and caching overhead of complex tag structures. 
Firstly, during the tag initialization phase, we designate seed nodes or edges for the query graph. 
A new tag is initialized only when a seed node or edge is matched. 
Secondly, during the tag propagation phase, we have designed an optimization algorithm for query graph matching, which helps to reduce the number of query graph nodes that need to be matched in each propagation.
We also implement a tag removal mechanism based on the time and the number of propagation rounds to further control the number of labels in the system. 
Additionally, we can split the query graph in a way that reduces the complexity of a single query.

\section{Realtime Graph Alignment with Tag Propagation Framework} 
\label{sec:tag_propagation}

\begin{figure}[t]
    \centering
    \includegraphics[width=0.48\textwidth]{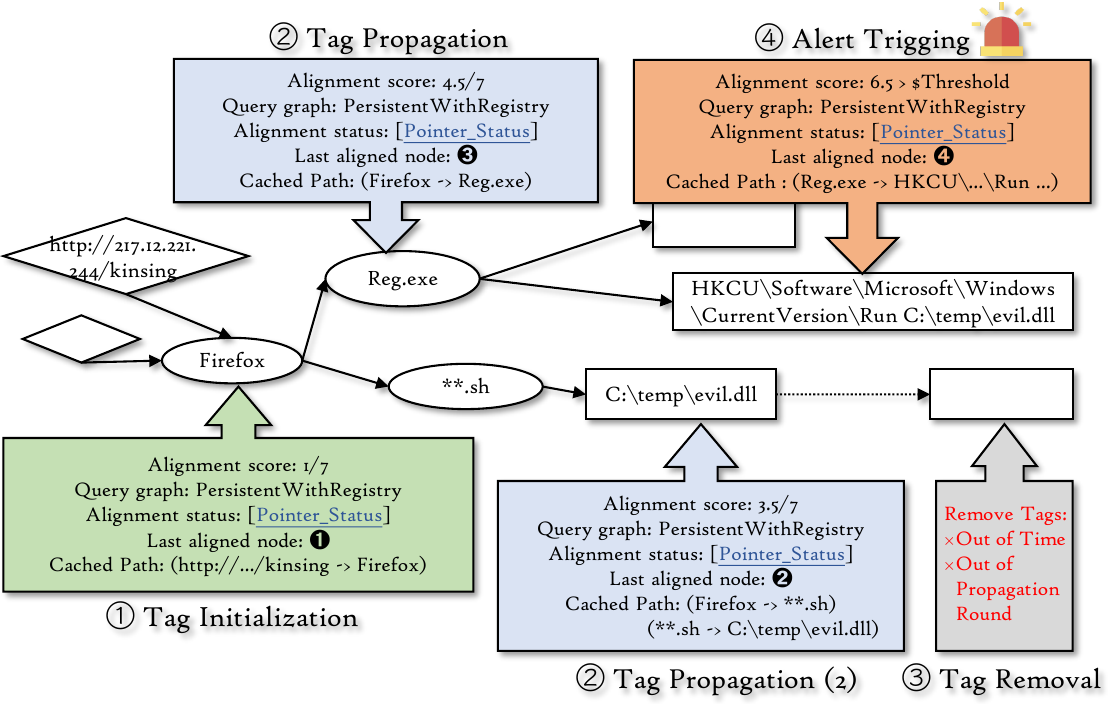}
    \vspace{-0.1in}
    \caption{Tag Propagation Process of Attack Technique ``\textit{T1547 - Autostart with Registry Run Keys}''}
    \vspace{-0.1in}
    \label{fig:tag_propagation}
\end{figure}

As a stream-based algorithm, tag propagation only handles one event at a time and thus reduces processing and caching overhead. 
The tag design is presented in Table~\ref{tab:tagcache}, and the overall detection process and the final result are presented in Fig.~\ref{fig:tag_propagation} and Table~\ref{tab:status} correspondingly.
As shown, the process consists of 5 major phases, including the tag lifecycle (initialization, propagation, and removal), tag merging, and alert trigging.
An example is presented in Fig.~\ref{fig:tag_propagation} for clarification.

\subsection{Tag Design}

Caching is a commonly used space-for-time trick in computer science.
Recently, several approaches~\cite{hossain2017sleuth, morse, conan} have adopted tags, which are attached to nodes in provenance graphs, as a cache of intermediate results for efficient provenance analysis. 
In this paper, we propose a new tag design tailored for streaming provenance graph alignment. 
In streaming graph alignment, each event can only be handled once. 
Thus, tags need to cache potentially aligned subgraphs (Node List, Edge List), as well as active fields (Active Node, Cached Path) and computational intermediate results (Aligned Score), as Table~\ref{tab:tagcache} shows.

\begin{table}[t]
\footnotesize
    \centering
    \caption{Key Fileds in Tags and Graph Alignment Status}
    \vspace{-0.1in}
    \begin{tabular}{c|c|c}
    \toprule 
         & Fields & Description \\
    \midrule 
        \multirow{5}{*}{Tag} 
                             & Query Graph (QG) & The query graph this tag try to align \\
                             & Align Score (S) & The overall alignment status as score \\
                             & Active Node ($N_{A}$) & Last query graph node aligned in tag \\
                             & Cached Path (Path) & Path from the last aligned node \\
                             & *\underline{Align Status} (Status) & Pointer to alignment status instance \\
    \midrule 
        \multirow{2}{*}{\makecell{Status}} & Node List & Aligned node list\\
                                         & Edge List & Aligned edge list\\
    \bottomrule
    \end{tabular}
    \vspace{-0.1in}
    \label{tab:tagcache}
\end{table}

\begin{table}[t]
\footnotesize
    \centering
    \caption{Graph Alignment Status of the Example}
    \vspace{-0.1in}
	\begin{tabular}{c|c|c|c}
	    \toprule {\makecell{Node\\/Edge ID}} & \makecell{Node/Edge\\Properties} & \makecell{Aligned Result} & \makecell{Align\\Score}\\
		\midrule 
            {\ding{202}} & [process] & firefox & -\\
		  {\ding{203}} & [executable] & C:$\backslash temp\backslash evil.dll$ & 1\\
		  {\ding{204}} & reg.exe & reg.exe & 1\\
		  {\ding{205}} & $hkcu\backslash$*$\backslash run\backslash$* & \makecell{$hkcu\backslash software\backslash microsoft\backslash w$\\$indows\backslash currentVersion\backslash run$} & 1\\
		\midrule 
            \ding{172} & {\ding{202}$\rightarrow$\ding{203}} & \makecell{(firefox $\rightarrow$ **.sh)\\(**.sh $\rightarrow C:\backslash temp\backslash evil.dll)$} & 0.5\\
		  \ding{173} & {\ding{202}$\rightarrow$\ding{204}} & (firefox $\rightarrow$ reg.exe) & 1\\
		  \ding{174} & {\ding{204}$\rightarrow$\ding{205}} & (reg.exe $\rightarrow$ $hkcu\backslash ...\backslash run ...$) & 1\\
            \midrule Total & \multicolumn{2}{c|}{T1547 - Autostart with Registry Run Keys} & 5.5/6\\
		\bottomrule
    \end{tabular}
    \vspace{-0.1in}
    \label{tab:status}
\end{table}

Among the fields, the Query Graph (QG) illustrates the corresponding query graphs for a tag.
The Align Score (S) is a numerical evaluation of the graph's overall alignment status. 
The Active Node ($N_A$) and Cached Path (Path) keep a record of the last aligned node and the subsequent path during the tag propagation process. 
These two active fields can be useful for speeding up graph alignment search, determining tag removal, and so on.
The Node List and Edge List archive complete information for additional graph analysis.

Furthermore, query graphs that represent attack behaviors often have various branches, as illustrated by the two "\ding{173} Tag Propagation" instances in Fig.~\ref{fig:tag_propagation}. 
It should be noted that the propagation of individual labels is unidirectional, making it impossible for the ancestor nodes to know if their following descendants are matched as well.
The tags on one branch also can not reveal the alignment status of the other sibling branches. 
As a solution, we implement a secondary caching mechanism (*\underline{Align Status}) to share alignment status among tags on different branches.
Therefore, all nodes in the same alignment would have the same alignment status.

\begin{algorithm}[ht]
    \small
    \caption{Tag Propagation-based Graph Alignment}\label{algo:tag_propagation}
    \SetAlgoLined
    \KwData{streaming event: $S_E=[e_i]$, query graph list: $L_{QG}=[QG_i]$}
    \KwResult{graph alignment Tags: $T_{GA}$}

    $M_{TC} \leftarrow$ \textbf{new} \textsc{Map}($\text{node}, T_{GA}$) \tcp{init tag cache}

    \ForEach{event $e_i=(N_{si}, R_i, N_{di})$ in $S_E$}{
    
        \tcc{\textbf{Tag Initialization}}
        $(QG_j,seed) \leftarrow$ \textsc{SearchSeed}($e_i, L_{QG}$)\\
        \If{$seedNode \neq null$}{
            $T_{GAJ} \leftarrow$ \textsc{InitTag}($QG_j, seed$) \\
            $M_{TC}.\textsc{Put}(N_{si}, T_{GAj})$
        }
        
        \If{$M_{TC}$.\textsc{Contains}($N_{si}$)}{
                
                \tcc{\textbf{Tag Removal}}
                \uIf{$T_{GA}'$.\textsc{IsOutDated}()}{
                    $M_{TC}$.\textsc{RemoveTag}($T_{GA}'$)
                }
                \ElseIf{$T_{GA}'$.\textsc{IsOutRounded}()}{
                    \Continue
                }
                
                \tcc{\textbf{Tag Propagation}}
                $T_{GA}' \leftarrow T_{GA}.\textsc{Get}(N_{si}).\textsc{PropagateTag}(e_i)$\\
            
                \tcc{\textbf{Tag Merging}}
                \eIf{$M_{TC}.\textsc{Contains}(N_{di})$}{
                    $T_{GA}' \leftarrow M_{TC}.\textsc{Get}(N_{di}).\textsc{MergeTag}(T_{GA}')$ 
                }{
                    $M_{TC}.\textsc{Put}(N_{di}, T_{GA}')$ 
                }  
                \tcc{\textbf{Alert Triging}}
                \If{$T_{GA}'$.\textsc{ShouldRaiseAlert}()}{
                    $M_{TC}$.\textsc{RemoveTag}($T_{GA}'$)\\
                    \Return{$T_{GA}'$}
                }
        }

    }
\end{algorithm}

\subsection{Tag Initialization}

The tag initialization, propagation, and removal lifecycle accounts for most of the overhead of \SysName. 
Limiting tag initialization can effectively reduce the system overhead.
Inspired by Poirot~\cite{milajerdi2019poirot}, we pick one or more seed nodes and events in each query graph. 
During the initialization phase of \SysName, the seeds of all query graphs are registered into a $(seed \rightarrow query\ graph)$ map.
During the stream processing phase, as lines 3-7 in Algorithm~\ref{algo:tag_propagation} show, tags will be initialized only when seeds are matched. 
Then, these initialized tags will be cached in $M_{TC}$, which is a $(node \rightarrow tag)$ map.

When selecting seeds for a query graph, it's important to choose the nodes and events at the early stages of the attack chain in the causal graph, which ensures that the entire attack can be captured during propagation. 
To minimize the number of tags, it's recommended to use unique nodes or events as seeds to initialize the query graph. 
If necessary, a multi-events subgraph alignment can be used as a condition to initialize the tag for whole graph matches for better overall performance.

\subsection{Tag Propagation}

Tags are created to store the results of the detection process. Specifically, in this paper, tags are used to store the partial graph alignment results, i.e., a set of suspicious behaviors. The tag propagation process involves two tasks: firstly, to transfer the tags and their states to the relevant nodes, and secondly, to search for nodes and edges that match the query graphs.

\begin{algorithm}
\small
    \caption{Tag Propagation (\textsc{PropagateTag}())}\label{algo:propagate_tag}
    \SetAlgoLined
    \KwData{graph alignment Tags: $T_{GA}$, event: $e_i$}
    \KwResult{new graph alignment Tags: $T_{GA}'$}

    $T_{GA}' \leftarrow$ \textbf{new} \textsc{GraphAlignmentTag}($T_{GA}$) \tcp{Copy constructions} 
    $T_{GA}'$.Path.add($e_i$)\\
    matchResult $\leftarrow T_{GA}$.QG.\textsc{Search}($N_{A}$, $e_i$)\\
    \If{matchResult $\neq \text{null}$}{
        $T_{GA}'.N_{A} \leftarrow N_{di}$ \\
        $T_{GA}'$.Path $\leftarrow$ \textbf{new} \textsc{Path}()\\
        $T_{GA}'$.S $\leftarrow T_{GA}'$.Status.\textsc{UpdateStatus}(matchResult, $T_{GA}'$.Path) \\
    }
    \Return{$T_{GA}'$}\;
\end{algorithm}


Lines 14-19 in Algorithm~\ref{algo:tag_propagation} and \ref{algo:propagate_tag} illustrate the tag propagation process. 
In this process, tags with partial alignment results search for subsequent alignments as they travel across the graph with the stream, completing overall graph alignment.
Specifically, for each income event ($e_i$), we will check if tags exist on its source node in $M_{TC}$.
If so, tags on source nodes, including the ones just initialized, will trigger the \textsc{PropagateTag}() function. 
A graph alignment tag records the graph alignment state up to the current node. 
Therefore, the first step in propagation is to copy the contents of existing tags ($T_{GA}$) to new instances ($T_{GA}'$) for new nodes. 
Meanwhile, $e_i$s will be added to the cached path in new tags.

We then check if the current event is one of the next steps in the query graph for the currently active node ($N_A$). 
The process can also be viewed as searching for the current event in a query graph ($GQ.\textsc{Search}(N_A,e_i)$), and the active nodes are set up to limit the scope of the query and reduce the overhead.
Specifically, for each propagation, the search in the query graph only covers neighboring nodes of the active node. This ensures a constant complexity search task irrespective of the query graph's size.
If the current event is found, we update the latest align to the tag and alignment status, as lines 5-7 shown in Algorithm~\ref{algo:propagate_tag}.

When propagating tags, it's possible for multiple tags to be propagated to the same node. If the tags correspond to different query graphs, we will keep all of them. However, if they belong to the same query graph, we need to merge them. To do so, we use a simple greedy strategy that aims to retain as many matching events as possible with high scores.


\subsection{Tag Removal}
\label{subsec: tag removal}

Dependency explosion is a universal problem that all provenance graph analysis encounters~\cite{li2021threat}. 
In our tag propagation scenario, dependency explosion leads to the overspreading of the tag, 
which could involve excessive computational overhead and ultimately make the solution unusable.
Therefore, except for limiting the initialization and propagation of tags, it is also necessary to set up tag removal mechanisms~\cite{hossain2020combating}.

We design two mechanisms for tag removal, namely, time-based (\textsc{IsOutDated}()) and propagation round-based (\textsc{IsOutRounded}()), as shown in lines 9-13 in Algorithm~\ref{algo:tag_propagation}, 
the time-based method removes tags based on their duration of existence, which can be implemented by setting expiration times for key-valued pairs in $M_{TC}$, while the propagation-based method stops tags propagation after a certain number of propagation rounds (\textsc{Length}(Path)). 
These values will be reset every time a new node is aligned or two tags are merged. 
By adjusting these thresholds, we can balance the depth and time window of detection with the overhead. 
More experimental results of the effectiveness of the tag removal mechanisms and threshold selection are discussed in \S\ref{subsec:RQ1}.

\subsection{Alert Triggering}

As line 7 in Algorithm~\ref{algo:propagate_tag} shows, the overall graph alignment score will be updated each time the tag propagates.
The score is defined by Equation (\ref{equ:edge_score}) and (\ref{equ:qg_score}):

\begin{equation}
\label{equ:edge_score}
\small
S_{ei}=S_{sink}+\frac{1}{path\_length}
\end{equation}

\begin{equation}
\label{equ:qg_score}
\small
S_{QG}=\frac{1}{2\times edge\_count}\times\sum _{i=0}^{edge\_count} S_{ei}
\end{equation}

Specifically, the alignment score for the whole graph ($S_{QG}$) is calculated by adding up the alignment scores for each of the edges ($S_{ei}$) in the graph and then dividing by twice the total number of edges ($edge\_count$).
The alignment score for each edge is the sum of the alignment score of its sink node ($S_{sink}$), which is fixed to 1 in this paper, plus the reciprocal of the path length($\frac{1}{path\_length}$).
The final alignment score ranges from 0 to 1; an example calculation is provided in Table~\ref{tab:status}.

Then, we can determine whether an alert should be triggered by comparing the score to a predefined threshold. 
Besides, it is worth noting that alerts generated by \SysName contain semantic information about nodes, paths, and attack types corresponding to query graphs, which can effectively assist subsequent attack analysis and response.

\section{Implementations}
\label{sec:implementation}

Based on streaming processing framework Flink~\cite{Flink} and Kafka~\cite{Kafka}, we implement \SysName with about 2K lines of Java code, which is publicly available online~\footnote{https://github.com}.
Specifically, we first read data from files and send original logs in string format to Kafka. 
Then, we adopt Flink, a stateful stream computation framework, to process the stream cached in Kafka.

A Flink streaming processing job can be divided into several operators of dependency, which means that the execution of the downstream operator(s) depends on the output produced by the upstream operator. 
These dependencies dictate the order and flow of data processing within the dataflow graph, ensuring that data is processed correctly and consistently.
Note that each operator can be subject to a separate parallelism setting, allowing flexible control over computational resources and enabling automatic scaling for larger-scale data synchronization.

In \SysName, the Flink job consists of four operators that work together. The first operator, called ``Kafka Source,'' reads string logs from Kafka. The second operator, ``Log Parsing,'' extracts attribute names and values from the strings. The third operator, ``Graph Construction,'' further extracts the attributes of interest and saves the node information needed for constructing normalized $(s,v,o,t)$ 4-tuple. Finally, the last operator, ``Graph Alignment,'' performs graph alignment on the 4-tuple stream.

\section{Evaluation}
\label{sec:eval}

This section systematically evaluated the efficiency, accuracy, and robustness of \SysName.
In particular, we investigate the following research questions (RQs):

\vspace{-\topsep}
\begin{list}{\labelitemi}{\leftmargin=1.5em}
 \setlength{\topmargin}{0pt}
 \setlength{\itemsep}{0em}
 \setlength{\parskip}{0pt}
 \setlength{\parsep}{0pt}
\item \textbf{RQ1:} How accurate is \SysName in detecting attacks? 
\item \textbf{RQ2:} How robust is \SysName against evasion attacks?
\item \textbf{RQ3:} How efficient is \SysName in handling streaming data? 
\item \textbf{RQ4:} How does scaling of host and query graphs affect \SysName's computation overhead?
\item \textbf{RQ5:} How does tag decay affect \SysName's performance?
\item \textbf{RQ6:} Are alerts generated by \SysName interpretable? (Appendix~\S\ref{appendix_rq6})
\end{list}

\subsection{Evaluation Setup}

\textbf{Dataset. } 
To evaluate the effectiveness of \SysName, we conducted experimental evaluations using two large-scale public datasets, comprising approximately 700 hours of log data and totaling nearly 250GB, with statistics information presented in Table~\ref{tab:dataset_efficiency}.

\begin{table*}[htbp]
    \footnotesize
    \centering
    \caption{Statistics of Logs and Processing Rate}
    \vspace{-0.1in}
    \begin{tabular}{c|c||c|c|c|c||c|c|c|c|c}
    \toprule 
        \multicolumn{2}{c||}{Datasets} & \makecell{\# of\\Entities} & \makecell{\# of\\Events}  & \makecell{Duartion\\(Data Size)} & \makecell{Avg. Event\\Generate Rate} & \makecell{\# of \\Init Tags} & \makecell{\# of \\Propagated Tags} & \makecell{ Processing\\Time} & \makecell{Avg. Event\\Process Rate} &  \makecell{Max\\Memory} \\
    \midrule 
        \multirow{3}{*}{\makecell{DARPA\\TC E3}}
                           & CADETS (C)   &626K  &15.2M &260h (35.63GB) &16eps &438k  &2,548k  &98s &155Keps  &464MB\\
                           & FiveDirections (F)  &239K  &20.6M  &172h (127.53GB) &33eps  &125K  &293K  &142s &145Keps  &364MB\\
                           & THEIA (T)  &1,142K  &8.1M  &135h (79.23GB) &17eps  &3,493K  &7,667K  &81s &100Keps  &623MB\\
    \midrule 
        \multirow{3}{*}{\makecell{PKU\\ASAL}} 
                           & WinServer (WS)  &11K  &8.3M  &32h (2.63GB) &72eps  &3K  &94K  &52s &159Keps  &206MB \\
                           & Win10 (W)  &66K  &7.9M  &30h (2.65GB) &73eps  &155K  &163K  &55s &144Keps  &245MB \\
                           & Linux (L)  &22K  &14.0M  &65h (9.75GB) &60eps  &123K  &431K  &113s &124Keps  &280MB \\
    \midrule 
        \multicolumn{2}{c||}{Overall}  &2,106K  &74.1M   &694h (257.42GB)  &45eps  &4,341K  &12,672K &541s &137Keps &363MB \\
    \bottomrule
    \end{tabular}
    \vspace{-0.1in}
    \label{tab:dataset_efficiency}
\end{table*}

{DARPA TC Dataset}~\cite{TC}.
DARPA Transparent Computing program conducted red team assessments from 2016 to 2019. Platforms with various services were set up, and benign activities were simulated. Engagement datasets contain millions of benign events and hundreds of malicious events, making detection a needle-in-a-haystack task. We select public Engagement 3 datasets from \THEIA, \CADETS, and Five Directions, collected on mainstream Linux, BSD, and Windows platforms for evaluation.

{PKU ASAL Dataset}~\cite{asal}.
The PKU ASAL dataset provided by NODLINK~\cite{li2023nodlink} consists of three hosts, namely, a Linux server running Ubuntu, a Windows server, and a Windows 10 host. These machines were used to carry out various normal activities, such as file downloads, execution of privileged commands, and compilation and execution of scripts.
Besides, this dataset comprises five emulated APT attack scenarios, which include APT29, Sidewinder, and Fin6. One attack was performed on the Linux server, three on the Windows host, and the final attack on the Windows server involved malicious activities like information gathering, privilege escalation, circumventing security tools, etc.



\textbf{Query Graphs } 
We constructed 12 distinct query graphs in total. Half of these graphs originated from MITRE ATT\&CK technique descriptions, while the remainder were derived from attack descriptions provided by the DARPA TC and PKU ASAL datasets. 
Query graph construction differs between attack techniques and complete attacks. For real-time detection, graphs for attack techniques require generality to detect variants, while those for complete attacks can prioritize specificity for accurate identification.
Characteristics of query graphs are illustrated in Table~\ref{tab:query-graph} and all query graphs are presented in Appendix~\S\ref{app:query_graphs}.

\textbf{Platform.} 
All experiments are performed on a laptop with AMD Ryzen 5 7600X 6-Core CPUs @ 4.70 GHz, 32 GB physical memory, and 1TB SSD, running Windows 11.

\begin{table}[htbp]
    \footnotesize
    \centering
    \vspace{-0.05in}   
    \caption{Characteristics and Aligned Results of Query Graphs}
    \vspace{-0.1in}
    \begin{tabular}{c|c|c|c|c|c}
    \toprule 
       \makecell{QG} & Query Graph Description  & \makecell{\# of \\ Vertexs} & \makecell{\# of \\ Edges} & \makecell{TPs \\ Graphs (Attacks)} & FPs\\
    \midrule
    1 & T1053 - Scheduled Task &4 &3 &13 (5) &0 \\
    2 & T1547 - Boot AutoStart &4 &3 &8 (8) &0 \\
    3 & T1003 - Credential Theft &5 &4 &1 (1) &0 \\
    4 & T1486 - Encrypt Data  &5 &4 &4 (2) &0 \\
    5 & Download\&Execution &7 &6 &13 (4) &0 \\
    6 & Live-off-the-Land &6 &5 &75 (5) &0\\
    \midrule
    7 & Nginx Backdoor 3.1 &6 &5 &1 (1) &0 \\
    8 & Nginx Backdoor 3.14 &8 &8 &1 (1) &0 \\
    9 & Browser Extension 3.10 &5 &4 &1 (1) &0 \\
    10 & Phishing E-mail 4.4 &4 &3 &1 (1) &1 \\
    11 & Browser Extension 3.3 &5 &4 &1 (1) &0 \\
    12 & Browser Extension 3.11 &10 &9 &1 (1) &0 \\
    \midrule
     \multicolumn{4}{c|}{Overall} & 120 (31) & 1 \\
    \bottomrule
    \end{tabular}
    \vspace{-0.05in}   
    \label{tab:query-graph}
\end{table}

\subsection{Evaluation Results}

\subsubsection{RQ1: How accurate is \SysName in detecting attacks? }
\label{subsec:RQ1}

To measure the accuracy of \SysName, we conducted graph alignment for all query graphs across the datasets. The results are detailed in Table~\ref{tab:query-graph}, with the alignment threshold set at 0.6.  
This threshold selection is further discussed in Appendix~\S\ref{appendix_threshold}. 
Overall, we aligned the query graphs with 121 subgraphs within the datasets, of which 120 were confirmed as true positives. This indicates that \SysName is highly precise in detecting threats that correspond to the query graphs. 
Besides, several query graphs successfully aligned with subgraphs in multiple datasets, demonstrating query graphs' generality in detecting attack variants in various scenarios.

Additionally, 31 of the aligned subgraphs have been confirmed as attacks in ground-truth reports. The remaining subgraphs represent potential malicious activities within the system. Without additional information, it is challenging to determine if these are actual attacks. Therefore, these will not be classified as either true positives or false positives in our accuracy calculations for attack detection. Instead, false positives are identified only in cases where tag over-propagation results in false alarms.

\begin{table}[htbp]
    \footnotesize
    \centering
    \vspace{-0.05in}   
    \caption{Comparison of Graph-level Detection Accuracy}
    \vspace{-0.1in}
    \begin{tabular}{c|c|c|c|c|c|c}
    \toprule
    \multirow{2}{*}{Dataset}  & \multicolumn{2}{c|}{Poirot} & \multicolumn{2}{c|}{ProvG-Searcher} & \multicolumn{2}{c}{\SysName} \\ \cmidrule(l){2-7}
                      & Precision &Recall &Precision  &Recall &Precision &Recall \\ \midrule
    CADETS             &0.99       &0.97          &0.99        &0.99        & \textbf{1.00}       & \textbf{1.00}  \\ 
    THEIA             &0.95      &0.99          &\textbf{0.99}        &0.99        & 0.92       & \textbf{1.00}  \\ 
    AVERAGE              &0.97         &0.98       &\textbf{0.99}        &0.99        & 0.96       & \textbf{1.00}   \\
    \bottomrule
    \end{tabular}
    \vspace{-0.05in}   
    \label{tab:comparsion}
\end{table}

\textbf{Comparsion with SOTA works.}
To further demonstrate the effectiveness of \SysName, we compared its performance in identifying attack graphs against two state-of-the-art graph-level detection approaches: Poirot ~\cite{milajerdi2019poirot} and ProvG-Searcher~\cite{provdetector}. The results, presented in Table~\ref{tab:comparsion}, are based on performance data from the ProvG-Searcher paper. 
As shown, although \SysName exhibits slightly lower precision than Poirot and ProvG-Searcher on the THEIA dataset, it outperforms these methods in both recall and precision in the other dataset.


\subsubsection{RQ2: How robust is \SysName against evasion attacks?}

\begin{table}[htbp]
    
\vspace{-0.1in}
    \footnotesize
    \centering
    \vspace{-0.05in}   
    \caption{Effective of Detecting Adversarial Mimicry Attack}
    \vspace{-0.1in}
    \begin{tabular}{c|c|c|c}
    \toprule 
     \makecell{Attack Cases \\ (Evading with in-chain insertion~\cite{mukherjee2023evading})}  & \makecell{ProvDetector}  & \makecell{NodLink}  & \makecell{\SysName} \\ \midrule
     \makecell{Powershell \\ (Invoke-AtomicTest T1003.003)}&\ding{55}   &\ding{55}  & \ding{51} \\ 
     7z (7z -tzip T1022.zip)  &\ding{55} &\ding{55}  &\ding{51} \\ 
     reg (reg add \#/Explorer/Run/)  &\ding{55}  &\ding{55}  &\ding{51} \\
    \bottomrule
    \end{tabular}
    \label{tab: mimicry attack}
    \vspace{-0.05in}   
\end{table}

Recent studies~\cite {mukherjee2023evading, goyal2023sometimes} demonstrate that mimicry attacks can effectively evade provenance-based intrusion detection systems. 
To evaluate the robustness of \SysName against these evasion tactics, we conducted three evasion attacks within the ASAL dataset. As shown in Table~\ref{tab: mimicry attack}, these attacks involved inserting normal events into attack paths to simulate system actions without altering the attack functionality. 

We then compared \SysName's performance against two other systems, ProvDetector and NodLink.
The results indicate that both ProvDetector and NodLink were successfully evaded after an average of 4-5 system actions were inserted. In contrast, \SysName consistently detected the attacks, demonstrating its superior robustness against evasion attempts.

\subsubsection{RQ3: How efficient is \SysName in handling massive data?}

To evaluate the efficiency of \SysName, we simultaneously processed 12 query graphs across all datasets. The results, summarized in Table~\ref{tab:dataset_efficiency}, indicate that \SysName processed a total of 257.42 GB of data, accumulated over 694 hours, in just 541 seconds. This corresponds to an average throughput of 137,000 events per second. Based on industry benchmarks, this rate is comparable to the data generation rate of approximately 6,000 PCs.

\begin{figure}[h]
    \centering
    \includegraphics[width=0.40\textwidth]{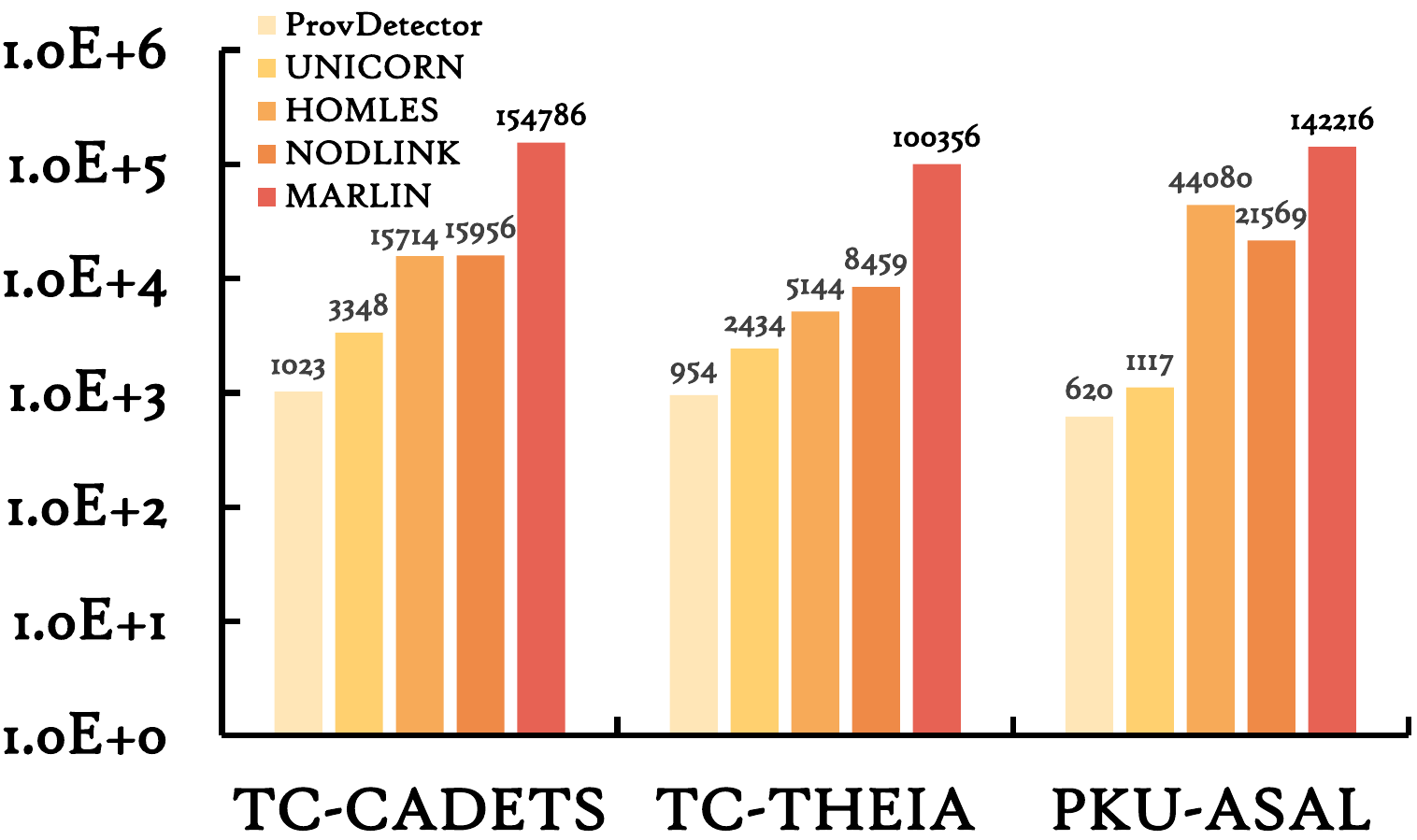}
    \vspace{-0.1in}
    \caption{Comparsion of Throughput with Existing Works}
    \vspace{-0.05in}
    \label{fig: comparison}
\end{figure}

\textbf{Comparsion with Existing Works.}
To further clarify the efficiency of \SysName, we compare it with several existing works of PIDSes, including NODLINK~\cite{li2023nodlink}, ProvDetector~\cite{provdetector}, UNICORN~\cite{han2020unicorn}, HOLMES~\cite{holmes}, and SAQL~\cite{gao2018saql}. 
Fig.~\ref{fig: comparison} illustrates a comparison of the throughput of \SysName and several other systems when processing the same datasets. 
The baseline result is from NODLINK~\cite{li2023nodlink}.
The experiments are conducted on the hardware with a 32-core CPU, which is much better than ours. 
Despite this, \SysName shows higher throughputs on every dataset. 
Therefore, we believe that \SysName is more efficient than these existing systems.
Besides, SAQL~\cite{gao2018saql} also shows a high throughput of up to 110K events per second on a 12-core CPU.
However, due to the lack of open-source systems and datasets, it is challenging to perform a fair comparison.
Nevertheless, the results from their original paper indicate that SAQL's memory occupation, approximately 2GB, and the overhead of parallel performing multiple queries are larger than ours.

\begin{figure}[htbp]
	\centering
	\begin{minipage}{0.49\linewidth}
		\centering
		\includegraphics[width=1.0\linewidth]{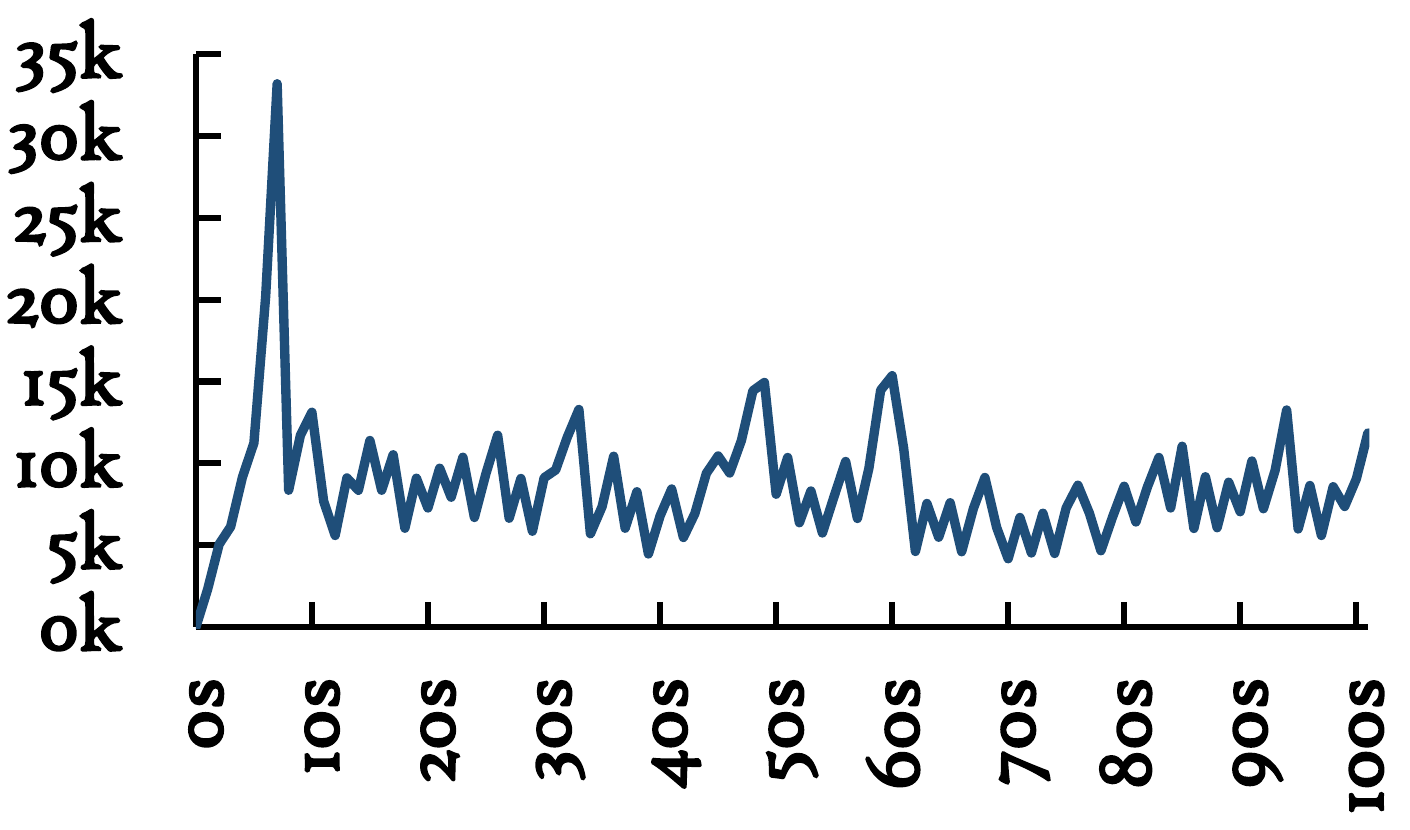}
            \textbf{(a) TC-CADTES}
	   \end{minipage}
	\begin{minipage}{0.49\linewidth}
		\centering
		\includegraphics[width=1.0\linewidth]{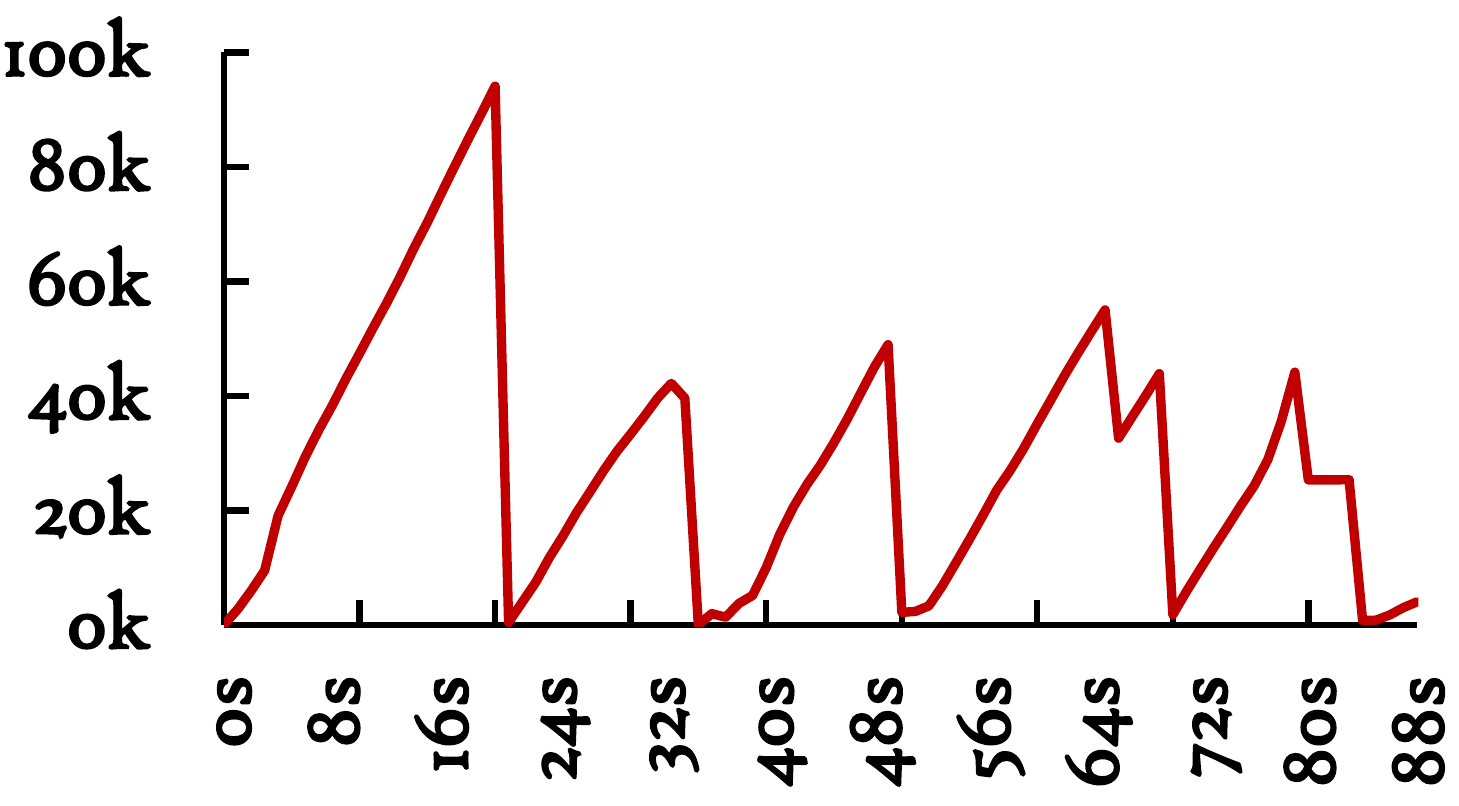}
            \textbf{(b) TC-THEIA}
	\end{minipage}

    \qquad

    \begin{minipage}{0.49\linewidth}
    \centering
    \includegraphics[width=1.0\linewidth]{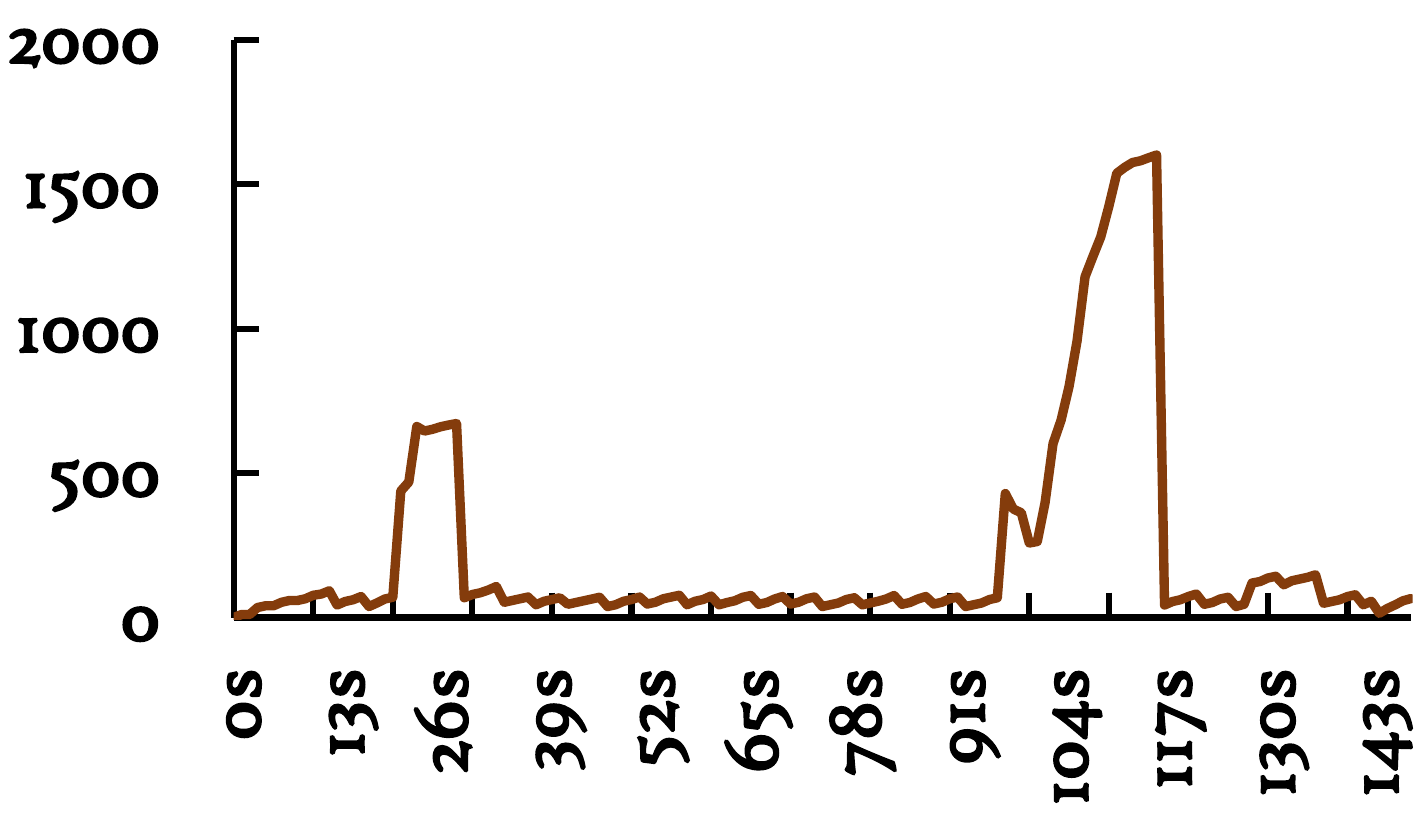}
    \textbf{(c) TC-FiveDirections}
	\end{minipage}
    \begin{minipage}{0.49\linewidth}
	\centering
	\includegraphics[width=1.0\linewidth]{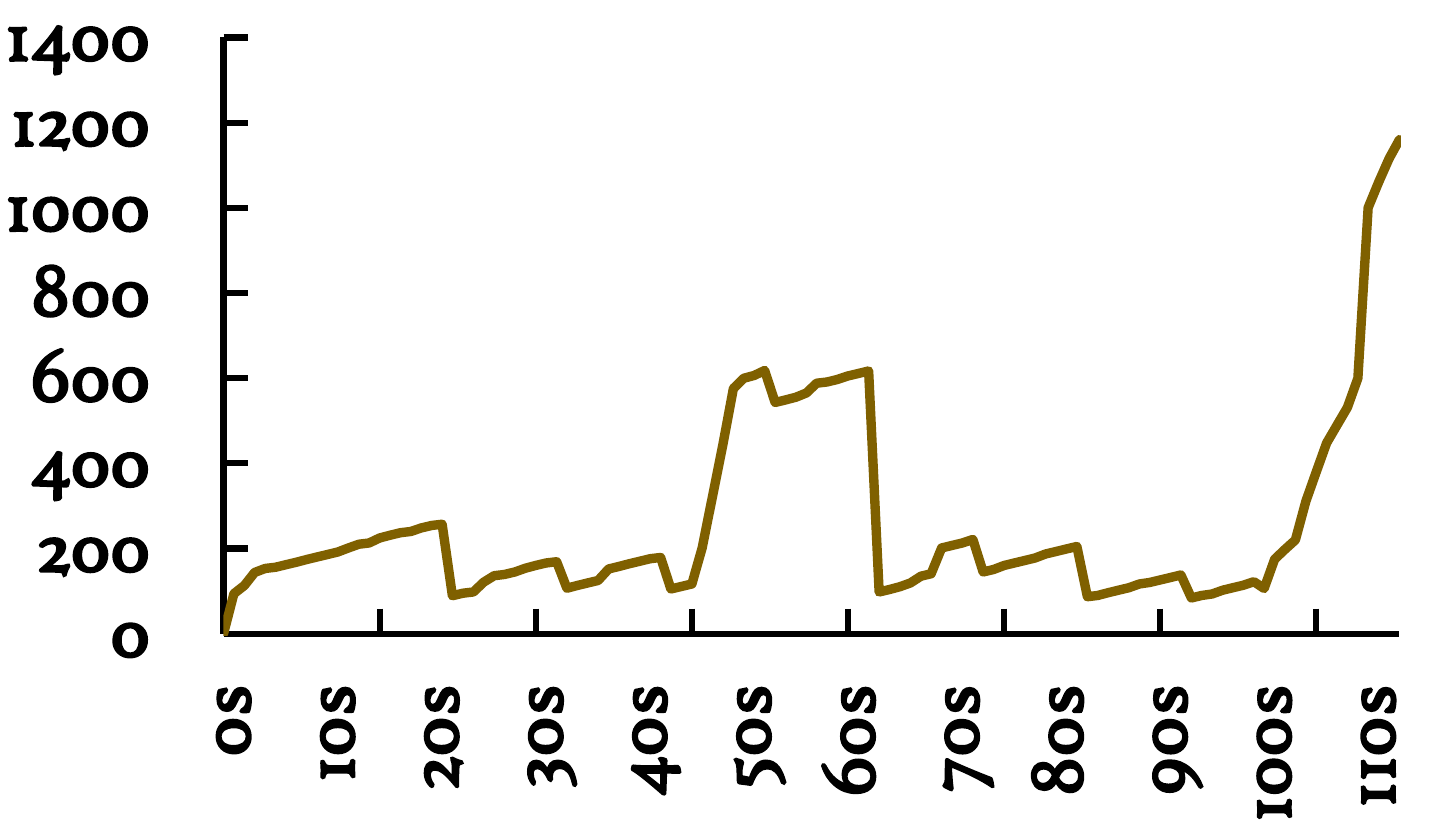}
        \textbf{(d) ASAL-Linux}
	\end{minipage}
    \vspace{-0.1in}
    \caption{The Number of Active Tags at Each Moment}
    \label{fig: combatting dependency explosion}
    \vspace{-0.05in}
\end{figure}

\textbf{Combatting Dependency Explosion. }
In large-scale provenance analysis, dependency explosion will impose unexcepted large overheads. Therefore, it is important for a provenance-based detection system to be able to control the dependency explosion, aka control the number of active tags in our scenario. 
In \SysName, tag decay time and rounds are setted to avoid infenitely propagation of tags.
Fig.~\ref{fig: combatting dependency explosion} counts the number of active tags at every moment when the system is running on different datasets.

The results demonstrate that the number of active tags fluctuates with several peaks over time but will not increase indefinitely. Peaks in tag counts occur when processing related events that contain numerous query graph seed nodes. Conversely, the number of active tags significantly declines during scheduled cleaning intervals, demonstrating effective control over dependency explosion.
Besides, due to the absence of timestamp attributes in the log data from ASAL's Windows Server and Windows 10 systems, these two datasets are not included in this experiment.

\begin{figure}[h]
    \centering
    \includegraphics[width=0.474\textwidth]{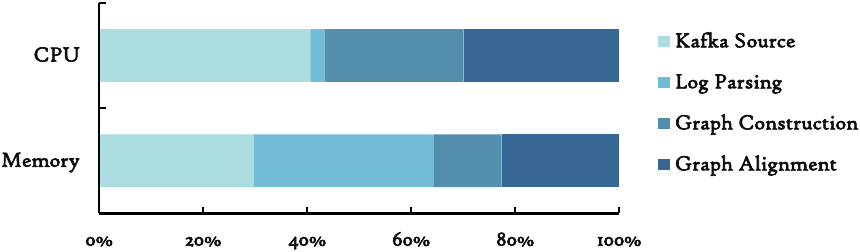}
    \vspace{-0.2in}
    \caption{CPU and Memory Occupation of Each Component}
    \vspace{-0.05in}
    \label{fig:occupation}
\end{figure}

\textbf{Breakdown Analysis. }
In order to analyze the computational resource usage, we decompose and examine the computational resource consumption of each Flink operator.
The task allocation for each operator has been discussed in \S\ref{sec:implementation}.
Brefily, the Kafka source operator pulls data from Kafka and the Flink framework. In contrast, the log parsing and graph construction operators parse raw string format logs and construct the streaming provenance graph, respectively.
The graph alignment operator aligns graphs with the tag-propagation framework. 
To analyze each module individually, we disabled one operator at a time, from the back to the front, and recorded the average time and memory consumption. 

The summarized results are presented in Fig.~\ref{fig:occupation}.
As shown, pulling data from Kafka requires most of the computational resources, accounting for approximately 40\% of the computation time and 25\% of the memory.
Besides, Raw log parsing occupies more memory as it caches some of the data for event merging.
The graph construction operator utilizes resources in a more balanced way. 
Finally, The graph alignment operator consumes only about 25\% of the overall resources. 
Therefore, if the data structure generated by the collector at the beginning meets the requirements, the detection overhead can be significantly reduced.

\subsubsection{RQ4: How does scaling of host and query graphs affect \SysName's computation overhead?}

In this section, we test \SysName's ability to process data from multiple hosts simultaneously and the impact of the number of query graphs on \SysName's performance.

\begin{figure}[htbp]
    \centering
    \vspace{-0.1in}
    \includegraphics[width=0.4\textwidth]{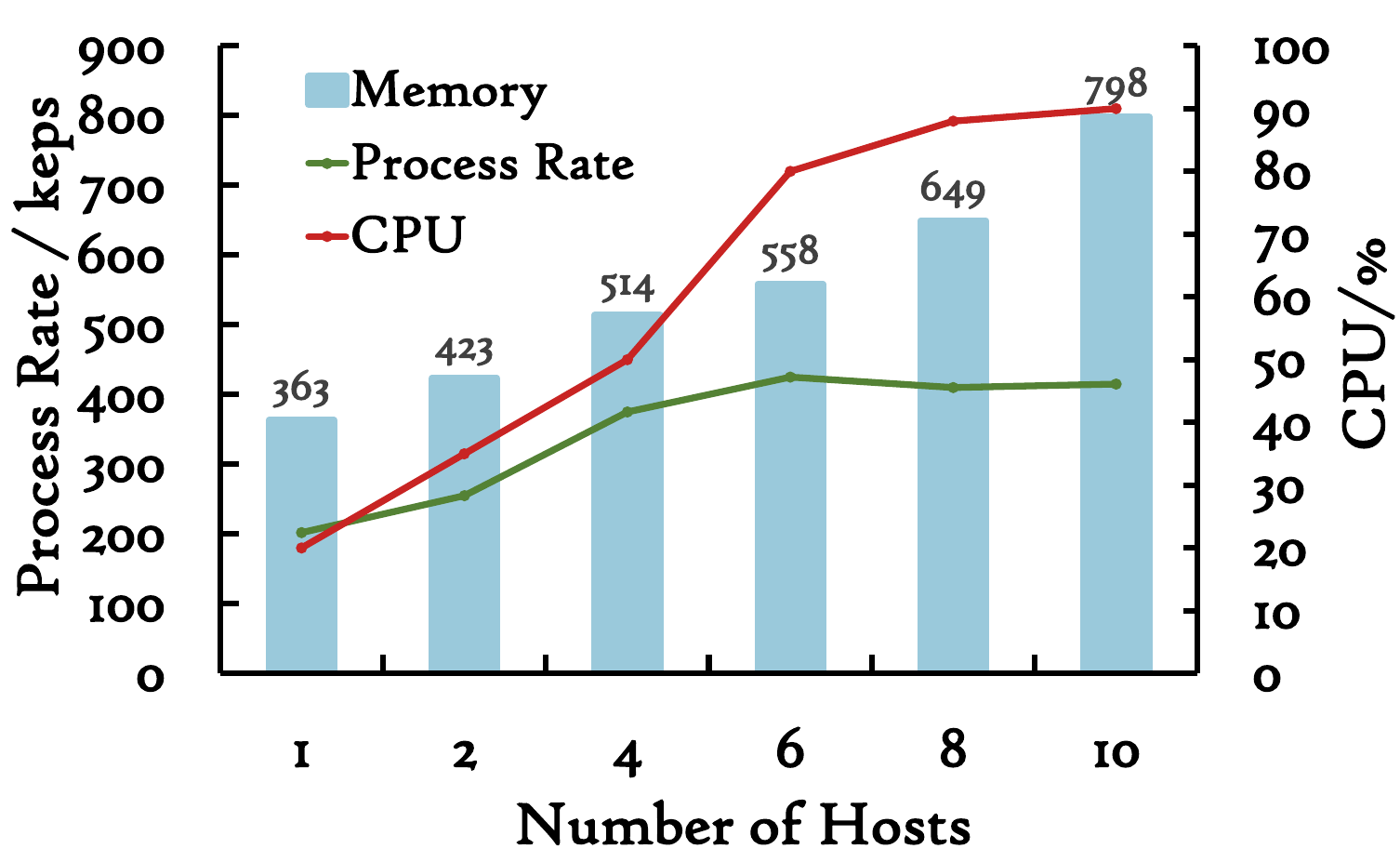}
    \vspace{-0.1in}
    \caption{Performance of Processing Multiple Hosts Log}
    \vspace{-0.1in}
    \label{fig: hosts}
\end{figure}

\textbf{Multiple Hosts. }
Table ~\ref{tab:dataset_efficiency} shows the performance of processing data from a single host, while \SysName's streaming framework can concurrently process log information from multiple hosts. To simulate the multiple-host scenario, we send log data to multiple Kafka topics and configure \SysName's hosts to consume data from these topics in parallel for concurrent processing. As the result shown in Fig.~\ref{fig: hosts}, the memory usage of \SysName increases with the increase in the number of topics, which is obvious. Similarly, when the number of Kafka topics increases from 1 to 6, the CPU utilization and processing rate increase nearly linearly. However, when the CPU utilization approaches 90\%, the growth of the processing rate decreases significantly. 
At this point, the system reaches its computational limit with no spare CPU capacity available. The increase in concurrent processes necessitates greater CPU resources for managing multi-threading, I/O, and memory allocation, which consequently slightly diminishes the processing rate.

\begin{figure}[ht]
    \centering
    \vspace{-0.1in}
    \includegraphics[width=0.4\textwidth]{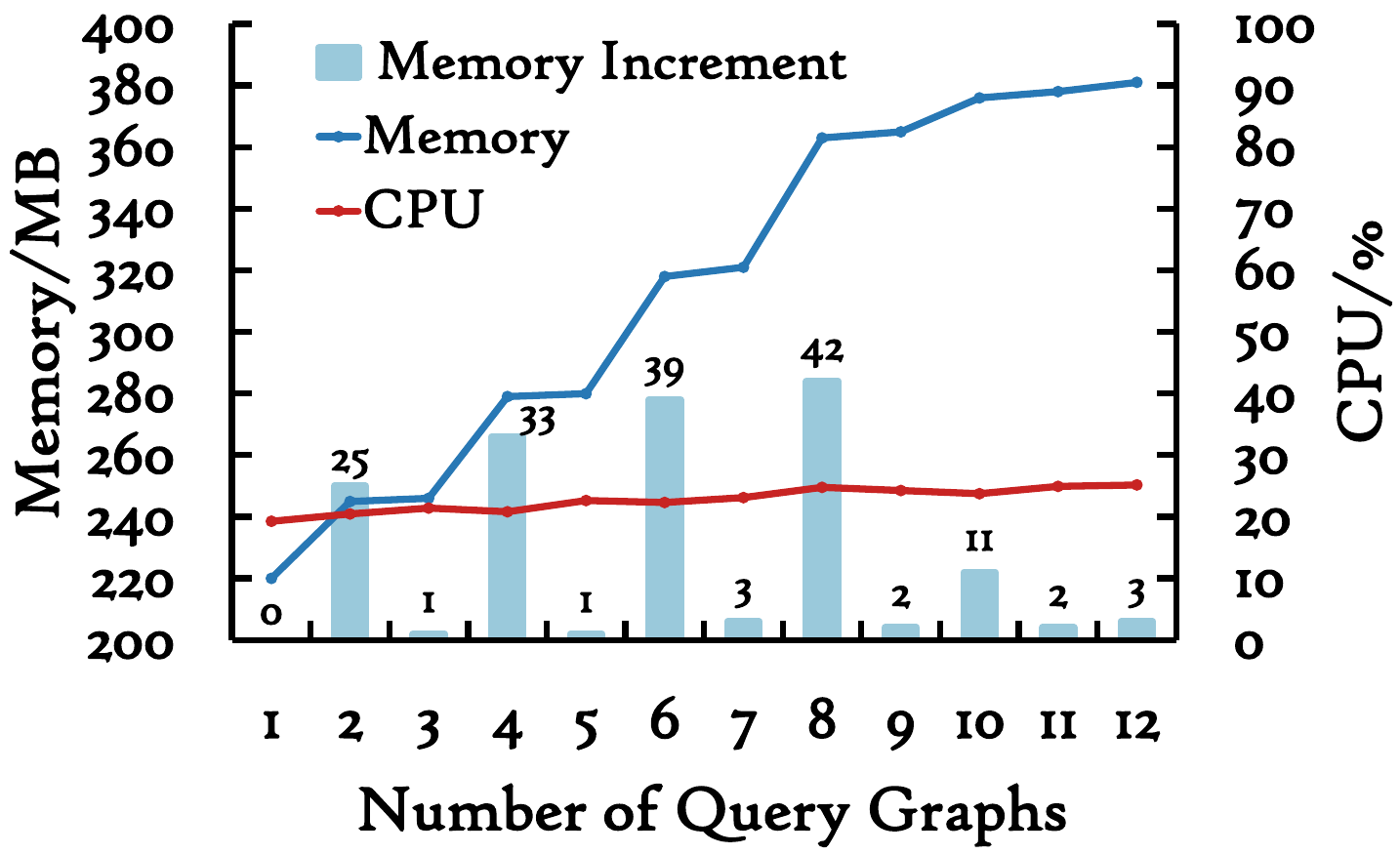}
    \vspace{-0.1in}
    \caption{Performance of Processing Multiple Query Graphs}
    \vspace{-0.1in}
    \label{fig: subgraphs}
\end{figure}

\textbf{Multiple Query Graphs}.
\SysName is designed to simultaneously match multiple attack graphs, unlike Poirot and ProvG-Searcher, which are limited to matching only a single graph at a time. This section evaluates how increasing the number of query graphs affects \SysName's performance. As depicted in Fig.~\ref{fig: subgraphs}, query graphs 2, 4, 6, and 8 correspond to behaviors that frequently occur within the system, which primarily drives the increase in \SysName's memory usage.
The addition of query graphs predominantly impacts the loading of these graphs and the determination of seed nodes during system initialization. These processes are relatively operation-light within the \SysName framework, hence the minimal change in CPU usage despite the increased number of query graphs.

\subsubsection{RQ5: How does tag decay affect \SysName's performance?}
This section further explores how tag decay rounds and tag decay time, influence the accuracy and performance of \SysName.

\begin{figure}[ht]
    \centering
    \vspace{-0.1in}
    \includegraphics[width=0.4\textwidth]{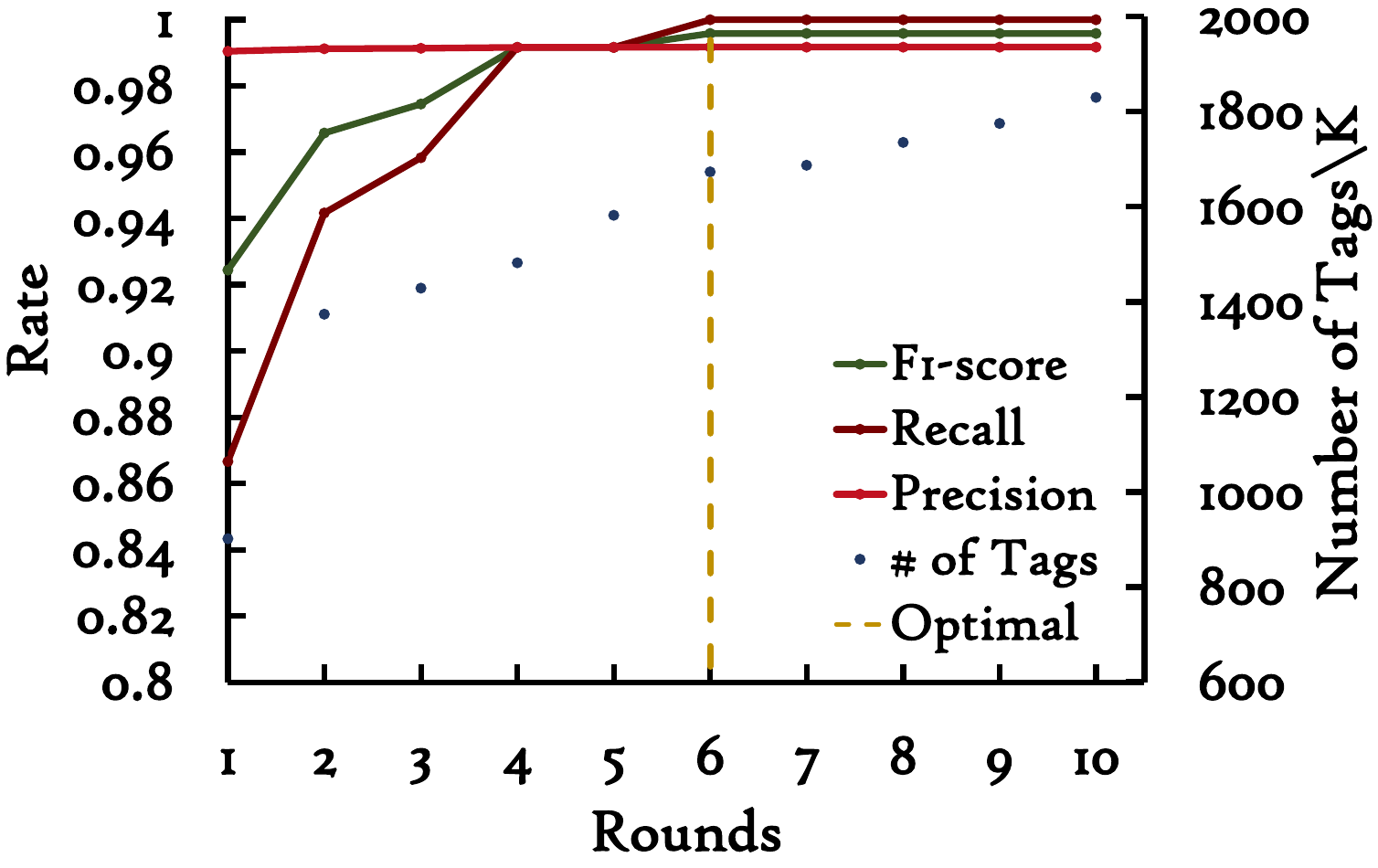}
    \vspace{-0.1in}
    \caption{Selecting the Optimal Tag Decay Rounds}
    \vspace{-0.1in}
    \label{fig: tag decay rounds}
\end{figure}

\textbf{Tag Propagation Rounds. }
The number of tag propagation rounds in \SysName directly influences the scope of node message passing, as illustrated in Fig.~\ref{fig: tag decay rounds}. When the number of rounds is set too low, the search range of \SysName narrows, resulting in decreased recall rates and reduced accuracy and robustness at the attack graph level. Conversely, setting the rounds too high leads to an excessive number of tags, which adversely affects detection performance due to increased memory overhead from the additional tags that must be cached.
Evaluation results indicate that setting the number of rounds to 6 allows \SysName to maintain high detection accuracy while balancing the memory overhead by controlling the number of tags generated during runtime across all datasets. 

\begin{figure}[ht]
    \centering
    \vspace{-0.1in}
    \includegraphics[width=0.4\textwidth]{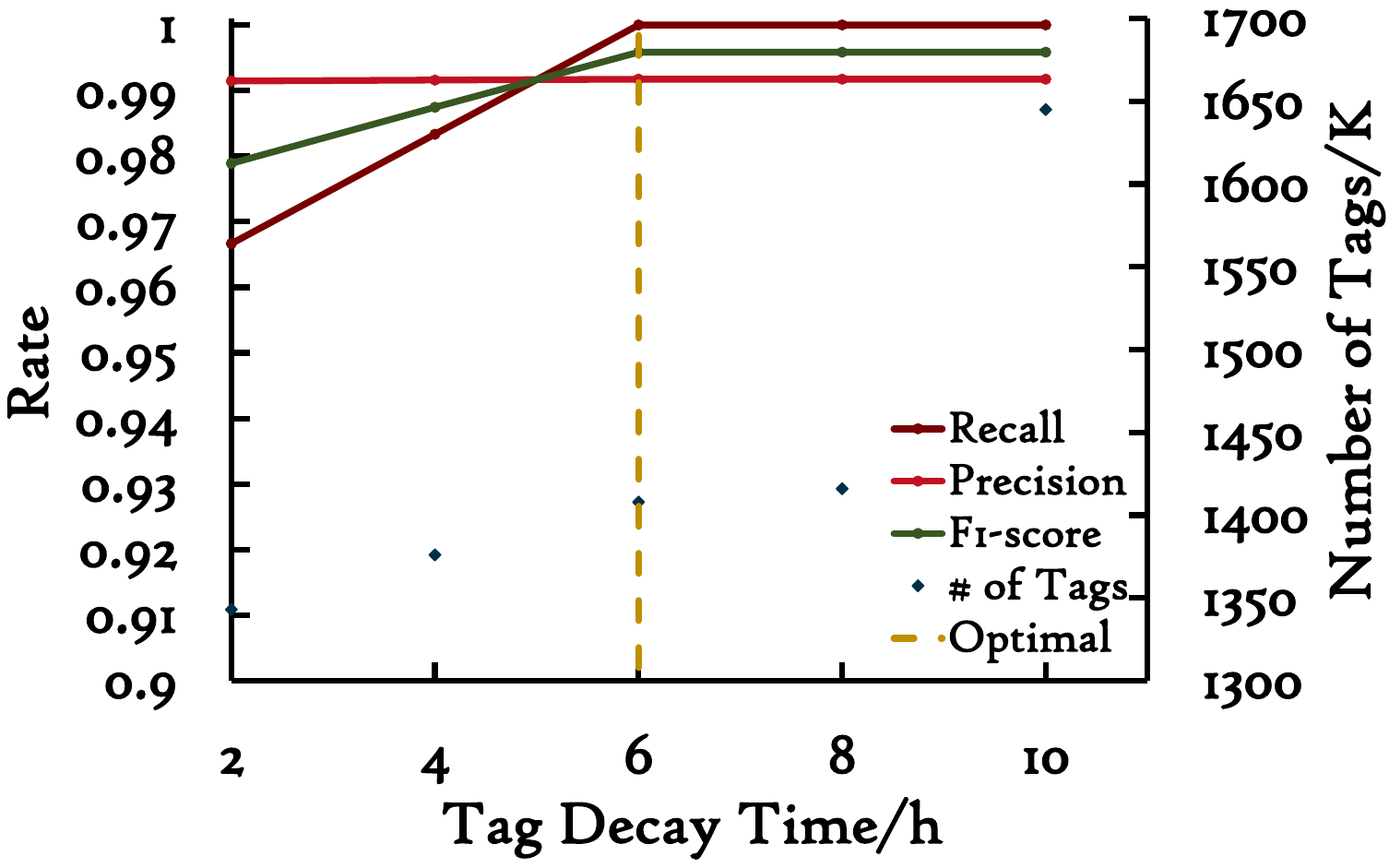}
    \vspace{-0.1in}
    \caption{Selecting the Optimal Tag Decay Time}
    \vspace{-0.1in}
    \label{fig: tag decay time }
\end{figure}

\textbf{Tag Decay Time. }
Similar to the constraints on tag propagation rounds, the length of tag decay time in \SysName is important, as it dictates how long the system retains memories of past suspicious events. Generally, a long decay time results in the accumulation of numerous tags. If these tags are not cleared promptly, \SysName's detection performance may suffer. Conversely, a very short decay time risks overlooking slow and stealthy attacks, as it might prematurely clear tags associated with such activities.

As illustrated in Fig.~\ref{fig: tag decay time }, the number of tags escalates rapidly when the decay time exceeds 8 hours. The growth in tag count is more gradual when the decay time is set between 2 to 8 hours. This moderate setting helps in efficiently clearing invalid tags without overly frequent cleanup, which is crucial as overly frequent tag clearance can compromise detection accuracy. Adversaries might conceal their activities over several hours, and frequent cleanups also consume valuable processing resources.

\section{Related Work}
\label{sec:related}

\subsection{Provenance-based Attack Detection}

In recent years, there's been growing interest in provenance-based detection for its effective intrusion detection capabilities. 

\textbf{Anomaly-based Approaches.}
NODOZE~\cite{nodoze} and its subsequent work, ProvDetector\cite{provdetector}, represent two state-of-the-art anomaly detection approaches for provenance data. NoDoze traverses the provenance graph to a specified depth and computes an anomaly score. While ProvDetector identifies rare paths within the graph and them into numerical vectors for anomaly score calculation. Such methods tend to produce more false positives and do not provide an interpretation of the results.

\textbf{Learning-based Approaches.}
Recent learning-based approaches, particularly those using GNNs, have demonstrated notable accuracy in cyberattack detection. For instance, ATLAS~\cite{alsaheel2021atlas} analyzes graph sequences using a Long Short-term Memory model for sequence classification. Similarly, UNICORN~\cite{han2020unicorn} processes histograms from provenance graphs, converting them into numerical vectors with the HistoSketch method~\cite{yang2017histosketch}. Meanwhile, SHADEWATCHER~\cite{zengy2022shadewatcher} employs a recommendation model to rank interactions by their likelihood of appearing in benign scenarios.
However, these learning class methods often incur significant computational overhead, restricting their embedding scope and making them susceptible to evasion tactics involving long attack chains.

\textbf{Tag-propagation-based Approaches.}
In tag propagation-based detection, event histories are condensed into tags and disseminated to capture subsequent interactions. These tags contain diverse information, such as numerical values representing confidentiality and integrity levels~\cite{hossain2017sleuth, morse}. Eventually, these tags aggregate to reach a predefined threshold, triggering alarms. For instance, SLEUTH~\cite{hossain2017sleuth} utilizes a tag propagation-based approach by defining two types of tags for each node, namely, trustworthiness tags and confidentiality tags. Similar tags design is also used in MORSE~\cite{morse} to emphasize critical aspects of the provenance graph. Conversely, HOLMES~\cite{holmes} constructs a high-level scenario graph (HSG) from the original provenance graph and maps scenarios to the lifecycle of APT attacks to determine alarm generation based on predefined policies.

\textbf{Mimicry Attacks.}
Mimicry attacks on Provenance-based Intrusion Detection Systems (PIDSes) have recently garnered significant attention. Goyal et al.\cite{goyal2023sometimes} suggest that adversaries can undermine detection capabilities by adding benign activities or structures to the provenance graph. Building on this, ProvNinja\cite{mukherjee2023evading} introduces the "gadget" chain concept, designed to make stealthy attacks less conspicuous. These strategies present new challenges for PIDSes.

\subsection{Continuous Graph Matching}

The incIsoMatch~\cite{fan2013incremental} algorithm, introduced by Fan et al. in 2013, was the first attempt to tackle the continuous subgraph matching (CSM) problem, but it suffered from low efficiency due to the enumeration of numerous outdated matches. Recently, the development of auxiliary structure-based approaches, such as GraphFlow~\cite{kankanamge2017graphflow} and RapidFlow~\cite{sun2022rapidflow}, has advanced CSM techniques with better performance. These algorithms primarily focus on conducting precise isomorphic matching. However, their matching definitions do not align with the need to detect various attack variants in provenance graphs, which require analyzing complex semantic information.

Besides, graph neural networks (GNN) and pre-training methods have been widely studied for graph matching. 
Models such as TGN~\cite{rossi2020temporal} and TGNF~\cite{song2021temporally} have advanced GNN techniques for learning Continuous Time Dynamic Graphs (CTDG), but challenges in optimizing storage and managing dynamically changing historical information persist.
In provenance graph analysis, POIROT~\cite{milajerdi2019poirot} introduces a graph alignment metric with a similarity-based search algorithm, which, while effective, demands substantial storage and only supports offline queries. To address these storage constraints, ProvG-Searcher~\cite{altinisik2023provg} implements a provenance graph partitioning strategy and a graph embedding algorithm to reduce storage demands. However, this approach struggles with the exponentially growing computational complexity associated with graph embedding.
\section{Discussion \& Conclusion}
\label{sec:discussion}

\subsection{Limitation}    

\textbf{Out of Observability.} 
Observability is crucial for effective intrusion detection. However, it's worth noting that some intrusion behaviors may not be detectable with basic provenance graphs. In some cases, lacking nodes or edges could lead to detection failure. To overcome this limitation, recent studies have proposed incorporating additional data layers, such as web~\cite{hassan2020omegalog} and microservices~\cite{datta2022alastor}, into the provenance graph. We can extend our detection capabilities by integrating these data into \SysName.

\textbf{Out of Query Graphs.} 
In this paper, we present a generalized framework for streaming graph alignment that can detect various types of attacks. 
The detection capability of the framework is determined by the query graph set used. 
It's important to note that attackers may design attack paths beyond the query graphs' scope despite their generalization ability.
To address this issue, analysts can create customized query graphs to flexibly extend \SysName's detection coverage.
Moreover, recent work on extracting graph-structured attack representations from cyber threat intelligence~\cite{li2022attackg} suggests that automating query graph extraction could be a promising direction for future work.

\subsection{Conclusion}

This paper proposes \SysName, a robust and efficient intrusion detection system that utilizes tag-propagation-based streaming provenance graph alignment for accurate attack detection. \SysName adopts attack knowledge embedded in query graphs and a tag-propagation framework, circumventing the substantial memory overhead associated with caching the entire provenance graph and the excessive computational burden typically encountered with conventional graph-matching algorithms.
Our experimental evaluations on two large-scale public datasets demonstrate that \SysName is efficient enough to process 176K events per second while accurately identifying 31 attacks and 3 evasions in massive data, significantly outperforming the state-of-the-art methods.

\bibliographystyle{ACM-Reference-Format}
\bibliography{AKG}

\appendix
\section{Mutated Attacks}
\label{app:mutation}

\begin{figure*}[htbp]
    \centering
    \includegraphics[width=0.96\textwidth]{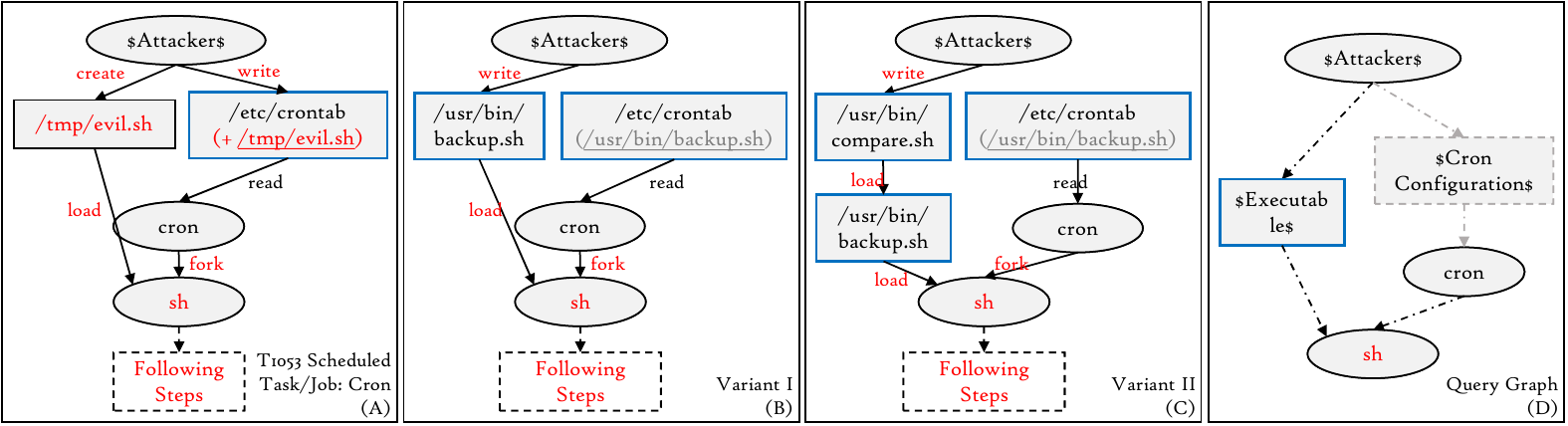}
    \vspace{-0.06in}
    \caption{An Example of Mutated Attack (Scheduled tasks (A) based on cron are commonly used threat persistence approaches. In some complex scenarios (BC), attackers can evade trivial rule-based detection by exploiting the existing cron tasks' control flow. However, utilizing a graph-based pattern (D), we can accurately identify such attack mutations.) }
    \vspace{-0.1in}
    \label{fig:motivating}
\end{figure*}

To further illustrate the effectiveness and efficiency of \SysName, we highlight the challenges associated with detecting an attack  mutations through a motivating example, depicted in Fig.~\ref{fig:motivating}.

Monitoring specific system entities or single-step behavior is a common threat detection strategy. 
For example, Fig.~\ref{fig:motivating} (A) demonstrates a simpler way for attackers to utilize cron for scheduled tasks and thus persistence (\textit{T1053 - Scheduled Task/job: Cron}~\footnote{https://attack.mitre.org/versions/v6/techniques/T1035/}). 
Defenders can easily detect such attacks by monitoring the cron configuration file.
However, as shown in Fig.~\ref{fig:motivating} (B) and (C), attackers can construct attack variants by hijacking the normal execution flow of a cron task. 
Specifically, in the first variant, the attacker rewrites the \texttt{/usr/bin/backup.sh} that is already in the cron scheduled task list to inject the attack logic.
Attackers can further exploit the execution flow to inject the attack logic to \texttt{/usr/bin/compare.sh} that execute by \texttt{backup.sh}, as the second variant shows.
Merely monitoring involved file changes could result in many false alarms as the execution chain gets longer.
Thus, we need a more general and semantic-aware attack behavior representation to cover these variants and keep a low false-positive rate.

Graphs are a highly expressive representation of cyber attacks and are effective in detecting attacks as a graph-matching or graph-querying problem~\cite{milajerdi2019poirot, li2022attackg}.
An example of a query graph corresponding to \textit{T1053} is presented in Fig.~\ref{fig:motivating} (D).
As shown, this query graph is general enough to contain the key behaviors for persistence with cron and match any of the above variants.
Instead of an exact match, we adopt a fuzzy graph matching (graph alignment) method to cover attack variants, including three fuzzy strategies: fuzzy node matching, fuzzy path matching, and incomplete graph matching.
Specifically, Nodes in a graph can be matched with a set of regular expressions based on their attributes, while edges in query graphs may correspond to multi-hop paths in the provenance graphs. 
Besides, it is not necessary to align the entire query graph with the provenance graph. Partial matching and reaching pre-defined thresholds are considered matches.

\section{Query Graphs}
\label{app:query_graphs}

This section gives 12 query graphs and their descriptions.


\begin{figure}[htbp]
    \footnotesize
	\centering
	\begin{minipage}{0.49\linewidth}
	   \centering
	   \includegraphics[width=1.0\linewidth]{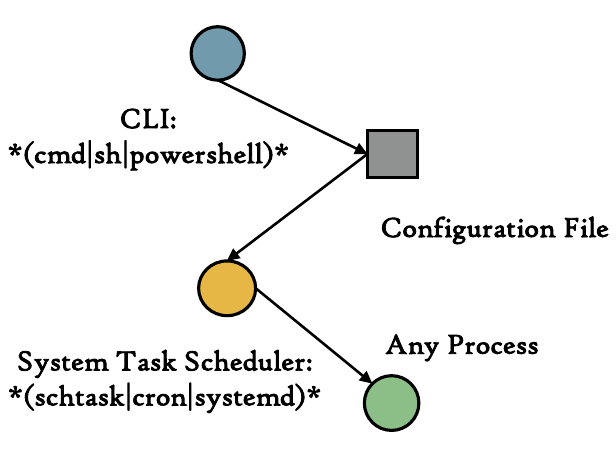}
          \textbf{T1053 - Scheduled Task}
	\end{minipage}
	\begin{minipage}{0.49\linewidth}
		\centering
		\includegraphics[width=1.0\linewidth]{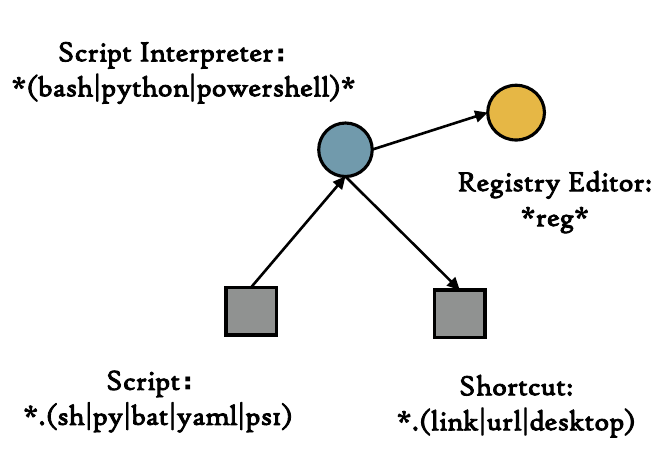}
        \textbf{T1547 - Boot AutoStart}
	\end{minipage}

    \footnotesize
	\centering
    \begin{minipage}{0.49\linewidth}
        \centering
        \includegraphics[width=1.0\linewidth]{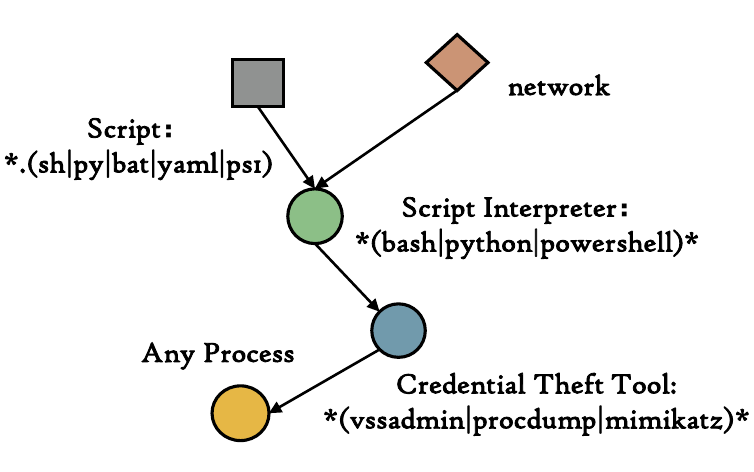}
        \textbf{T1003 - Credential Theft}
	\end{minipage}
    \begin{minipage}{0.49\linewidth}
	   \centering
	   \includegraphics[width=1.0\linewidth]{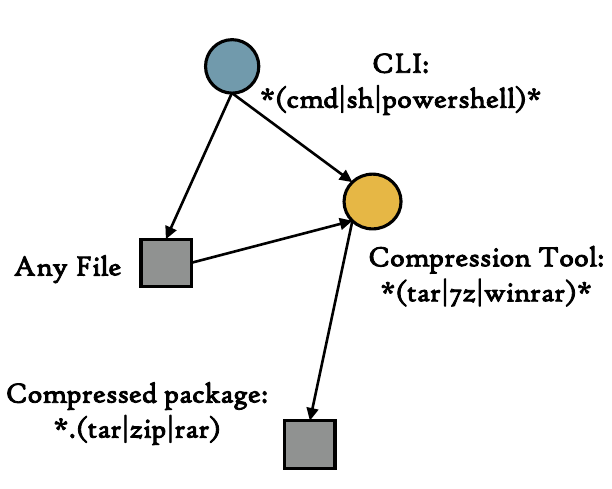}
          \textbf{T1486 - Encrypt Data}
	\end{minipage}

    \footnotesize
	\centering     
    \begin{minipage}{0.49\linewidth}
        \centering
        \includegraphics[width=1.0\linewidth]{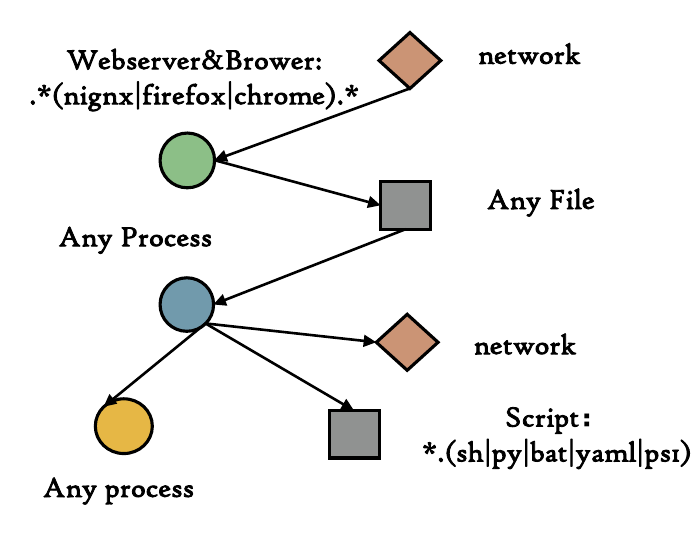}
        \textbf{Download\&Execution}
	\end{minipage}
    \begin{minipage}{0.49\linewidth}
	   \centering
	   \includegraphics[width=1.0\linewidth]{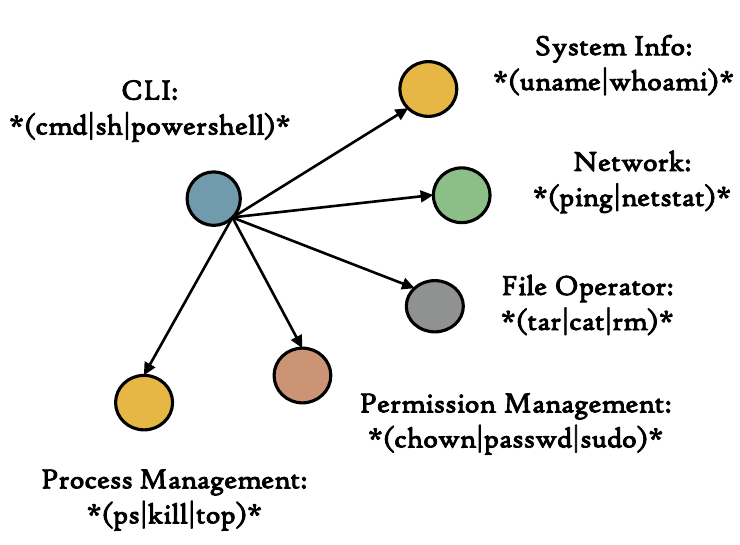}
          \textbf{Live-off-the-Land}
	\end{minipage}
 \caption{Technique-level Query Graphs}
\end{figure}

\begin{figure}[htbp]
    \footnotesize
	\centering
    \begin{minipage}{0.49\linewidth}
        \centering
        \includegraphics[width=1.0\linewidth]{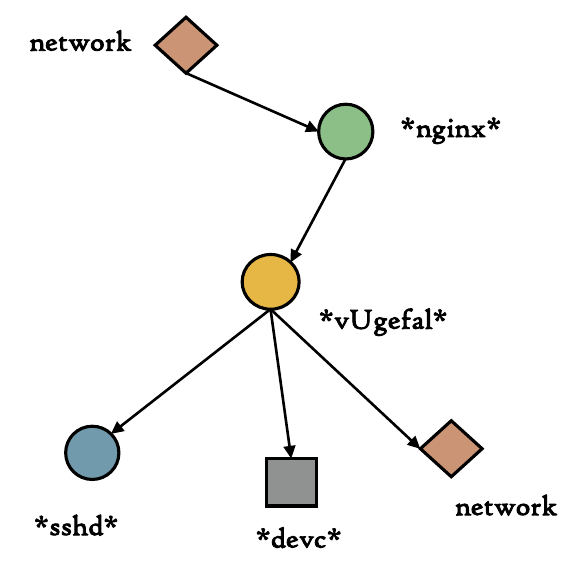}
        \textbf{Nginx Backdoor 3.1}
	\end{minipage}
    \begin{minipage}{0.49\linewidth}
	   \centering
	   \includegraphics[width=1.0\linewidth]{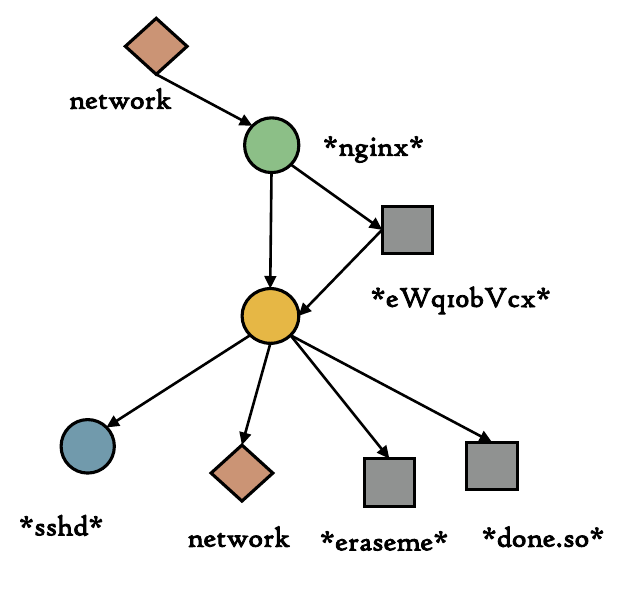}
          \textbf{Nginx Backdoor 3.14}
	\end{minipage}

    \footnotesize
	\centering     
     \begin{minipage}{0.49\linewidth}
        \centering
        \includegraphics[width=0.8\linewidth]{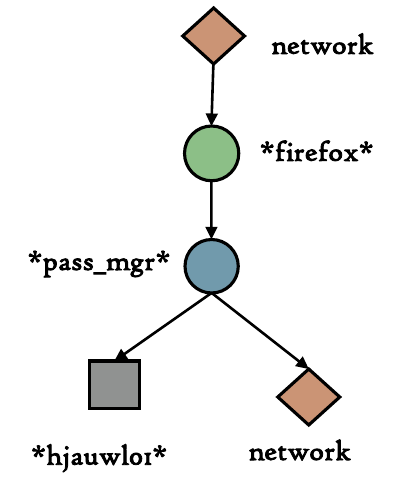}
        \textbf{Browser Extension 3.10}
	\end{minipage}
    \begin{minipage}{0.49\linewidth}
    	\centering
    	\includegraphics[width=1.0\linewidth]{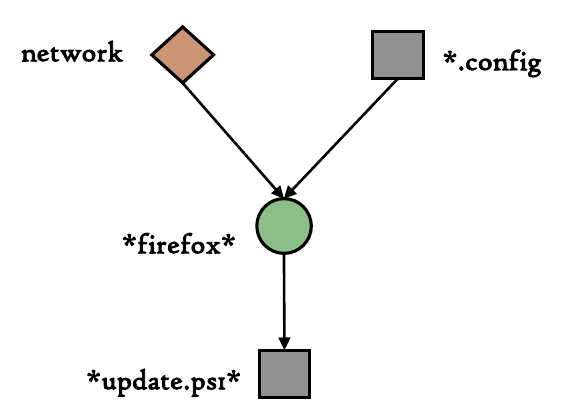}
            \textbf{Phishing E-mail 4.4}
	\end{minipage}

    \footnotesize
	\centering
     \begin{minipage}{0.49\linewidth}
         \centering
         \includegraphics[width=1.0\linewidth]{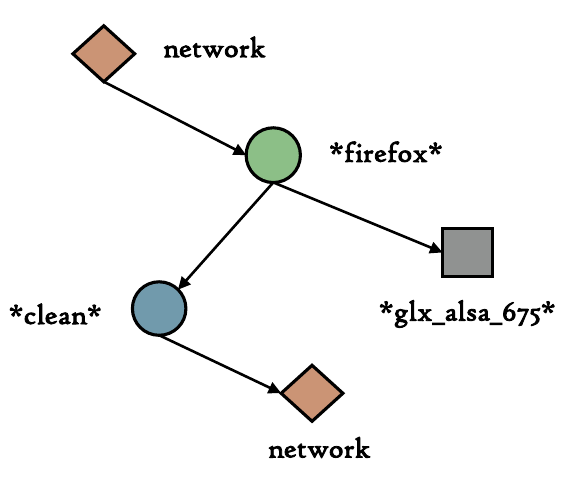}
         \textbf{Browser Extension 3.3}
	\end{minipage}
    \begin{minipage}{0.49\linewidth}
    	\centering
    	\includegraphics[width=1.0\linewidth]{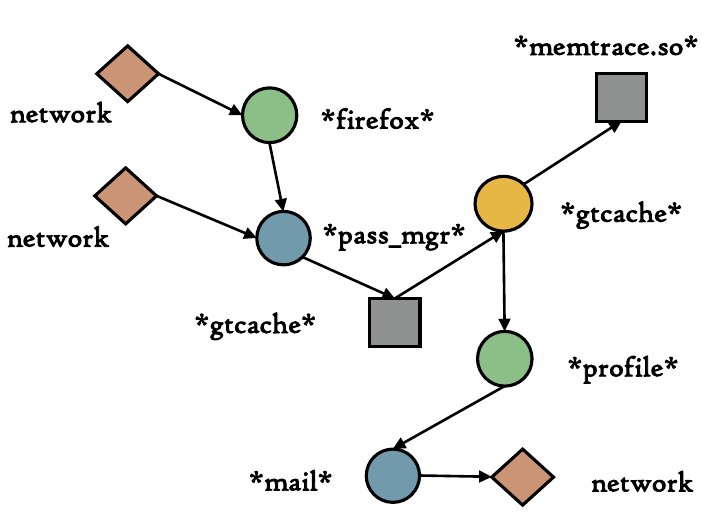}
        \textbf{Browser Extension 3.11}
	\end{minipage}
 \caption{Attack-level Query Graphs}
 \vspace{-0.1in}
\end{figure}

\textbf{(QG1) T1053 - Scheduled Task:} Adversaries may abuse the system service like \texttt{cron} to perform task scheduling by modifying configuration

\textbf{(QG2) T1547 - Boot AutoStart:} Adversaries can achieve persistence by adding programs to executable shortcuts or modifying the registry run keys

\textbf{(QG3) T1003 - Credential Theft:} Adversaries utilize credential theft tools to steal information

\textbf{(QG4) T1486 - Encrypt Data:} Adversaries encrypt data on target systems by utilizing encrypt tools.

\textbf{(QG5) Download\&Execution:} Adversaries can utilize webServer \& Browser to download a file, which is read by other processes to execute system behaviors.

\textbf{(QG6) Live-off-the-Land:} Adversaries utilize \texttt{cmdline} interface of the target system to perform tasks.

\textbf{(QG7) Nginx Backdoor 3.1:} Adversaries attack CADETS by exploiting vulnerabilities in Nginx to gain access to the CADETS FreeBSD system.

\textbf{(QG8) Nginx Backdoor 3.14:} Adversaries re-exploit Nginx with an HTTP request and implanted a new \texttt{drakon} malware in Nginx's memory, which established a connection with the C2 operator console.

\textbf{(QG9) Browser Extension 3.10:} Adversaries attack against FiveDirections by trying to exploit the target via the malicious pass manager browser extension in Firefox 54.0.1.

\textbf{(QG10) Phishing E-mail 4.4:} Adversaries download the update.ps1 script and used common threat attacks such as email phishing and Excel spreadsheet macros to target FiveDirections.

\textbf{(QG11) Browser Extension 3.3:} Adversaries implant \texttt{drakon} malware into the Firefox process by exploiting vulnerabilities in Firefox 54.0.1, establishing a connection with the operator console, and carrying out continuous attacks.

\textbf{(QG12) Browser Extension 3.11:} Adversaries attack Firefox 54.0.1 on THEIA through a malicious Passport Manager browser extension, implanting \texttt{drakon}'s executable files into the target disk through the extension, posing a continuous threat.

\section{Additional experiments}

\subsection{Selecting the Alignment Threshold}
\label{appendix_threshold}

\begin{figure}[htbp]
    \centering
    \includegraphics[width=0.42\textwidth]{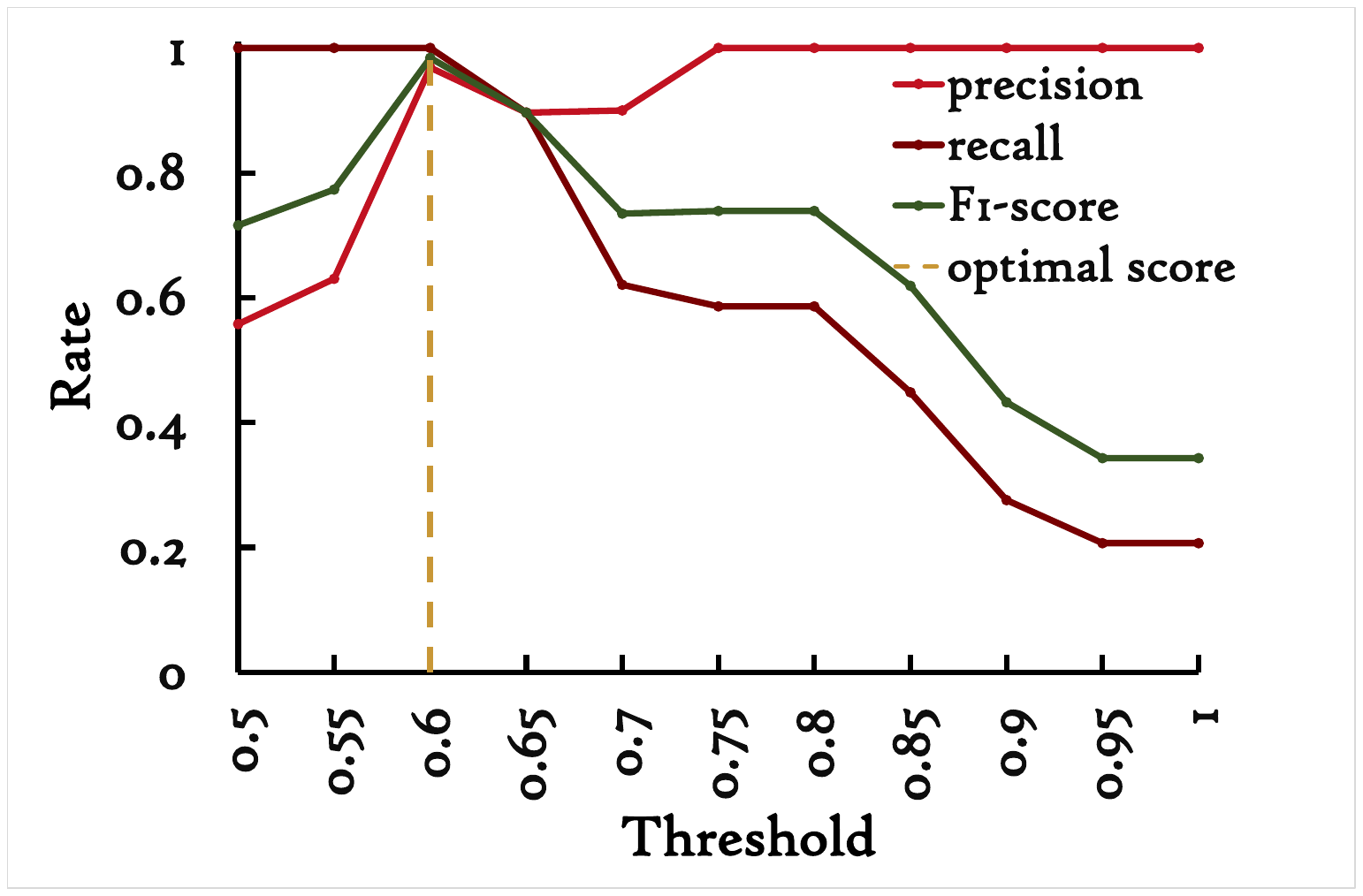}
    \vspace{-0.1in}
    \caption{Selecting the Optimal Threshold}
    \label{fig: threshold}
\end{figure}   

The selection of the threshold significantly influences the outcomes of detection. If set too low, the threshold may cause a surge in false positives, whereas an excessively high threshold could result in overlooking true positives. 
To identify the optimal threshold, we computed the $precision= \frac{TP}{TP+FP}$, $recall=\frac{TP}{TP+FN}$, and $F1\ Score=\frac{2TP}{2TP+FP+FN}$ across various threshold values using all available datasets, as depicted in Fig.~\ref{fig: threshold}. 
It's important to note that we manually verified the alignment of graphs at a low threshold (0.5) to establish ground truth for the aforementioned value computations.

The analysis results underscore the challenge of striking a delicate balance between detection accuracy and minimizing false alarms when employing a unified threshold selection method. In our context, prioritizing high detection rates takes precedence over minimizing false alarms.
Consequently, we have elected to adopt a unified threshold of 0.6, effectively capturing all true positives while concurrently minimizing false positives. Furthermore, analysts are encouraged to tailor thresholds to individual query graphs to optimize accuracy in practical applications.

\subsection{RQ6: Are alerts generated by \SysName interpretable? }
\label{appendix_rq6}

\begin{figure}[htbp]
    \centering
    \includegraphics[width=0.5\textwidth]{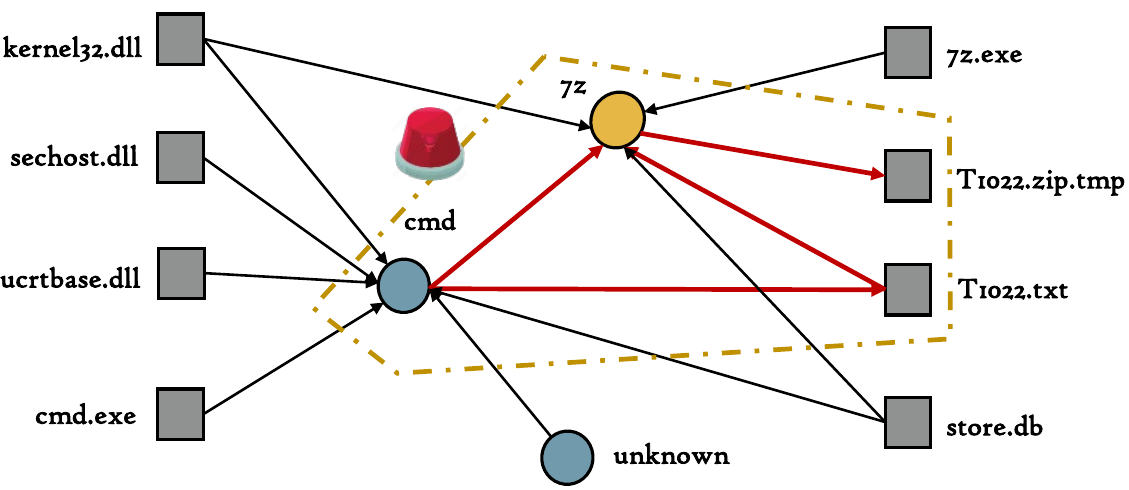}
    \vspace{-0.1in}
    \caption{A matched attack graph}
    \label{fig: example} 
\end{figure}

When designing MARLIN, we take the ability to reconstruct complete and concise attack processes as an important indicator, which can promote faster attack investigations for security analysts, especially given the lack of interpretability in many deep learning detection systems.
This interpretable description of attacks establishes trust in decisions among security analysts and accelerates the process of identifying and reducing false positives (FPs).

Fig.~\ref{fig: example} depicts the attack related to number \texttt{T1486} in the ASAL dataset, which is generally consistent with the ground truth of the dataset. The \texttt{unknown} process achieves file encryption operations by controlling \texttt{cmd} and \texttt{7z}, and the subgraph in the yellow dashed box matches the query graph. MARLIN will output this part of the content into the alarm information.
As shown in Table~\ref{tab:status}, during the tag propagation process, \SysName records the nodes and edges corresponding to the query graph.
This graph information containing precise attack behaviors can help analysts analyze alarms in time without the need to repeatedly query the original logs.

\textbf{Feasibility in Practice.}
In practice, intrusion detection systems are deployed in many continuously operating machines, which generate an enormous amount of logs every moment. 
Storing the raw logs, even for a small period of time, takes up a significant amount of memory. 
Short caching times result in the system being easily bypassed by long, stealthy APT attacks, while long caching times can lead to long response times and massive overhead. 
On the other hand, selective data caching can create new opportunities for attackers to bypass the system. 
Therefore, we choose a streaming scheme for efficient processing and responding.

Once the security operations center receives alerts, it is necessary to analyze and respond to the alarms manually.
However, false alarms and incomprehensible results can make analysis tasks complex and time-consuming.
To address this issue, we use graph-based detection to provide a graph-based description of detection results.
We also implement two strategies to reduce the false alarms and mitigate the dependency explosion problem, including limited tag initialization and removing tags based on the number of propagation rounds and time.

\end{document}